\documentclass[
reprint,
nofootinbib,
amsmath, 
amssymb,
aps,
]{revtex4-2}

\usepackage{graphicx}
\usepackage{dcolumn}
\usepackage{bm}
\usepackage{tabularray}
\usepackage{url} 

\usepackage{orcidlink} 

\newcommand\ba{\begin{eqnarray}}
\newcommand\ea{\end{eqnarray}}
\newcommand{\sv}{\langle \sigma v\rangle}

\newcommand{\eq}[1]{\begin{alignat}{2} #1 \end{alignat}}

\newcommand{\cog}[1]{{\boldmath #1}}
\newcommand{\cor}[1]{{#1}}

\newcommand{\cms}{{\rm cm}^3/{\rm s}}
\newcommand{\ee}{e^+e^-}
\newcommand{\gamgam}{\gamma\gamma}
\newcommand{\DM}{{\rm DM}}

\usepackage{subfig, caption}
\DeclareCaptionJustification{justified}{\leftskip=0pt \rightskip=0pt \parfillskip=0pt plus 1fil}
\usepackage{hyperref}
\hypersetup{
	colorlinks=true,
	linkcolor=blue,
	filecolor=magneta,      
	urlcolor=blue,
}
\begin{document}

\preprint{APS/123-QED}

\title[DM annihilation in EoR II]{Sensitivity toward dark matter annihilation imprints on 21-cm signal with SKA-Low: A convolutional neural network approach
}
%
%
\author{Pravin Kumar Natwariya\,\orcidlink{0000-0001-9072-8430}}
\email{pvn.sps@gmail.com}
\affiliation{School of Fundamental Physics and Mathematical Sciences, Hangzhou
Institute for Advanced Study, University of Chinese Academy of Sciences (HIAS-UCAS), Hangzhou, 310 024, China}
\affiliation{University of Chinese Academy of Sciences, Beijing, 100 190, China}
\affiliation{International Centre for Theoretical Physics Asia-Pacific (ICTP-AP), Beijing, 100 190, China}
\author{Kenji Kadota\,\orcidlink{0000-0003-2019-6007}}
\email{kadota@ucas.ac.cn}
\affiliation{School of Fundamental Physics and Mathematical Sciences, Hangzhou Institute for Advanced Study, University of Chinese Academy of Sciences (HIAS-UCAS), Hangzhou, 310 024, China}
\affiliation{University of Chinese Academy of Sciences, Beijing, 100 190, China}
\affiliation{International Centre for Theoretical Physics Asia-Pacific (ICTP-AP), Beijing, 100 190, China}
\author{Atsushi J. Nishizawa\,\orcidlink{0000-0002-6109-2397}}
\email{atsushi.nisizawa@gifu.shotoku.ac.jp}
\affiliation{Gifu Shotoku Gakuen University,
Takakuwanishi, Yanaizu, Gifu, 501-6194, Gifu, Japan \\
Institute for Advanced Research, and Kobayashi Maskawa Institute, Nagoya University,
Furosho, Chikusaku, Nagoya, 464-8602, Aichi, Japan}

\date{\today}

\begin{abstract}
This study investigates the sensitivity of the radio interferometers to identify imprints of spatially inhomogeneous dark matter annihilation signatures in the 21-cm signal during the pre-reionization era. We focus on the upcoming low-mode survey of the Square Kilometre Array (SKA-Low) telescope. Using convolutional neural networks (CNNs), we analyze simulated 3D 21-cm differential brightness temperature maps generated via the {\tt DM21cm} code, which is based on {\tt 21cmFAST} and {\tt DarkHistory}, to distinguish between spatially homogeneous and inhomogeneous energy injection/deposition scenarios arising from dark matter annihilation. The inhomogeneous case accounts for local dark matter density contrasts and gas properties, such as thermal and ionization states, while the homogeneous model assumes uniform energy deposition. Our study focuses on two primary annihilation channels to electron-positron pairs \((e^+e^-)\) and photons \((\gamgam)\), exploring dark matter masses from 1 MeV to 100 MeV and a range of annihilation cross-sections. For \(\gamgam\) channel, the distinction across dark matter models is less pronounced due to the larger mean free path of the emitted photons, resulting in a more uniform energy deposition. For \(e^+e^-\) channel, the results indicate that the CNNs can effectively differentiate between the inhomogeneous and homogeneous cases. Despite observational challenges, the results demonstrate that these effects remain detectable even after incorporating noise from next-generation radio interferometers, such as the SKA. We find that the inhomogeneous dark matter annihilation models can leave measurable imprints on the 21-cm signal maps distinguishable from the homogeneous scenarios for the dark matter masses $m_\DM=1$ MeV and the annihilation cross-sections of $\sv \geq 5 \times 10^{-30}~\cms$ ($\sv \geq 5 \times 10^{-29}~\cms$ for $m_{\DM}=100$ MeV) for moderate SKA-Low noise. Our results demonstrate the potential of machine learning techniques, such as CNNs, to extract subtle features from complex astrophysical data and sensitivity to probe exotic sources of energy injections, offering a novel approach to exploring the nature of dark matter and the early Universe with future 21-cm observations.
\end{abstract}

\maketitle
\section{Introduction}
\label{sec:intro}

The upcoming observations of the redshifted 21-cm signal, arising from the hyperfine transition of neutral hydrogen, are expected to provide insight into the era when the first luminous objects formed at an unprecedented level. According to the prevailing model of cosmology, the Lambda Cold Dark Matter ($\Lambda$CDM) model, dark matter plays a crucial role in the formation of structures in the Universe. Planck measurements indicate that dark matter constitutes approximately 27 percent of the total energy content of the Universe, far exceeding the 5 percent attributed to ordinary baryonic matter \cite{Planck:2018}. Despite overwhelming evidence for its existence--- from galactic rotation curves, gravitational lensing, {gamma-ray observations,} and the large-scale structure of the Universe--- its fundamental nature remains one of the most significant unsolved problems in astrophysics and cosmology {\cite{Planck:2018, Boddy:2022knd, Cooley:2022ufh, Feng:2010, Totani:2025fxx}}.

The observations such as the Experiment to Detect the Global Epoch of Reionization Signature (EDGES)\footnote{\href{https://www.haystack.mit.edu/astronomy/astronomy-projects/edges-experiment-to-detect-the-global-eor-signature/}{https://www.haystack.mit.edu/}} \cite{Monsalve:2016xbk}, Probing Radio Intensity at high-Z from Marion (PRI\(^Z\)M) \cite{PRIZM:2019}, Shaped Antenna measurement of the background RAdio Spectrum (SARAS)\footnote{\url{https://wwws.rri.res.in/DISTORTION/saras.html}} \cite{SARAS3:2021}, Broadband Instrument for Global Hydrogen Reionization Signal (BIGHORNS) \cite{BIGHORNS:2015}, Dark Ages Radio Explorer (DARE)\footnote{\url{https://lunar.colorado.edu/dare/}} \cite{DARE:2015, DARE:2017}, Large Aperture Experiment to Detect the Dark Ages (LEDA)\footnote{\url{http://www.tauceti.caltech.edu/leda/}} \cite{Spinelli:2022xra}, and Radio Experiment for the Analysis of Cosmic Hydrogen (REACH)\footnote{\url{https://www.reachtelescope.org/}} \cite{REACH:2022} aim to measure the monopole global redshifted 21-cm signal from the pre, post and during reionization era. 
However, such monopole measurements are intrinsically limited in their ability to constrain new physics, due to strong degeneracies among astrophysical parameters and their lack of sensitivity to spatial structure in the 21-cm signal.
The experiments that can observe the spatial variations of the 21-cm signal across different redshifts will provide a way to explore the presence of any new physics--- a combination of spatial and temporal information will be an additional advantage to resolve degeneracies.
The radio interferometers like the upgraded Giant Metrewave Radio Telescope (uGMRT)\footnote{\url{https://www.gmrt.ncra.tifr.res.in/}} \cite{uGMRT:2017}, Hydrogen Epoch of Reionization Array (HERA)\footnote{\url{https://reionization.org/}} \cite{HERA:2020, Hera:2021, HERA:2022}, Murchison Widefield Array (MWA)\footnote{\url{https://www.mwatelescope.org/}} \cite{MWA:2012}, Square Kilometre Array (SKA)\footnote{\url{https://www.skao.int/}} \cite{SKA:2019}, Low Frequency Array (LOFAR)\footnote{\url{https://www.astron.nl/telescopes/lofar/}} \cite{LOFAR:2003, LOFAR:2013}, and Precision Array for Probing the Epoch of Reionization (PAPER) \cite{PAPER:2013}, are designed to detect spatial variations in the redshifted 21-cm signal, enabling detailed three-dimensional mapping of the distribution of neutral hydrogen and ionized regions throughout all three important astrophysical phases that shaped the present-day Universe: Cosmic Dawn, Reionization, and Post-Reionization. 

The new generation of interferometer SKA Low-frequency (SKA-Low)\footnote{\url{https://www.skao.int/en/explore/telescopes/ska-low}} is particularly promising. It will provide direct images of the 21-cm signal over a broad redshift range ($z\sim$27 to $z\sim$3), covering all three critical phases \cite{Koopmans:2015}. The substantial improvement in sensitivity not only opens up new opportunities for groundbreaking science but also presents challenges, requiring the development of innovative analysis techniques to manage large, complex datasets. The present work explores the imprints of inhomogeneous dark matter annihilation in which energy injection varies spatially and how to distinguish it from scenarios involving homogeneous energy injection in the plasma. 
We use machine learning techniques, specifically convolutional neural networks (CNNs), to differentiate between homogeneous and inhomogeneous dark matter annihilation models using simulated three-dimensional 21-cm differential brightness temperature maps from the pre-reionization era. We add the system temperature noise associated with the SKA-Low mode survey in the simulated 21-cm differential brightness temperature maps. Understanding the impact of the inhomogeneities on the 21-cm signal maps can play a vital role in making accurate predictions in the dark matter paradigm.

Previous analyses of the 21-cm signal have primarily focused on summary statistics, with a particular emphasis on the power spectrum. While these methods have proven effective, they may not capture all the information encoded in the three-dimensional brightness temperature fields. These limitations become especially important when distinguishing subtle differences between various dark matter models or energy injection mechanisms into the plasma. Machine learning, as a rapidly advancing field, offers significant potential for revealing complex patterns within high-dimensional data that are difficult to detect by traditional statistics \cite{Murakami:2020, Sabiu:2022, Murakami:2024, Murakami::2024, Zhu:2019, Cheng:2021, Wu:2023, Gebhard:2019, Li:2020, Xia:2021, Baltus:2021, Zhang:2022, Caron:2018, Khek:2022, Khosa:2020}. In the context of 21-cm line cosmology, machine learning techniques, especially deep learning methods like CNNs, provide a novel method for image-based analysis \cite{Gillet:2019, Villanueva:2021, Sabiu:2022, Murakami::2024, Nishizawa:2024bnh}. It can automatically learn hierarchical features and identify complex spatial patterns that arise due to the presence of any new physics and subsequent effects on structure formations and astrophysical processes. This approach can emerge as a treasure trove for detecting the subtle effects of the annihilating dark matter models that produce only small deviations from the homogeneous energy injection mechanisms. 

We employ a modified version of the publicly available code {\tt DM21cm}\footnote{\url{https://github.com/yitiansun/DM21cm}} to generate realistic 21-cm signal maps \cite{Sun:2023}. The code uses semi-numerical cosmological simulation code {\tt 21cmFAST}\footnote{\url{https://github.com/joshwfoster/21cmFAST}}$^,$\footnote{\href{https://homepage.sns.it/mesinger/DexM___21cmFAST.html}{https://homepage.sns.it/mesinger/DexM\underline{\hspace{0.5cm}}21cmFAST.html}}  to produce three-dimensional cosmological realizations of different physical fields, such as the gas temperature, ionization fraction, spin temperature, 21-cm differential brightness temperature, etc., below a redshift of $z < 45$ \cite{21CMFAST:2011, Munoz:2022, Qin:2020, Park:2019, Murray:2020}.  For the temperature and ionization initial conditions required by {\tt 21cmFAST}, the {\tt DM21cm} code utilizes the {\tt DarkHistory}\footnote{\url{https://github.com/hongwanliu/DarkHistory/tree/DM21cm}} code \cite{Liu:2023fgu, Liu:2023nct, Liu_2020}. These codes help us efficiently explore the various dark matter masses and annihilation cross-sections. We produce brightness temperature maps on $64^3$ grids with a box size of 48 Mpc, providing sufficient resolution to capture relevant spatial fluctuations while maintaining computational feasibility.

The structure of the paper is as follows: In section \ref{21cmIntro}, we briefly discuss the 21-cm signal and power spectrum. Section \ref{ssec:energyinjection} describes how the spatial distribution of dark matter injects the energy into the surrounding medium and subsequent effects on the temperature and ionization of gas and Ly-$\alpha$ coupling. In section \ref{sec:ML}, we outline the machine learning methods used in the paper and their training and fine-tuning. Finally, section \ref{sec:result} presents the results obtained, and section \ref{sec:summary} summarizes the results. Throughout this study, we adopt Planck 2018 cosmological parameters \cite{Planck:2018}.

\section{Simulations and Initial Conditions}
\label{sec:simulation}

\subsection{21-cm Differential Brightness Temperature}\label{21cmIntro}

In this subsection, we discuss the 21-cm differential brightness temperature, $\Delta T_b$--- which is the observable quantity and can be used as a tracer of inhomogeneous dark matter annihilation. We consider the epochs before and during the onset of reionization. These epochs can be observed in the form of absorption or emission by the neutral hydrogen medium relative to the cosmic microwave background radiation (CMBR)--- spectrum temperature $T_{\rm CMB}$, or background radio radiation--- spectrum temperature $T_{\rm R}$, at a reference frequency of $\nu_0=1420.4$~MHz. The corresponding wavelength for this frequency is 21-cm. This originates from the hyperfine transition between the 1S singlet and triplet states of the neutral hydrogen atom. To study the state of the plasma at a redshift of $z$ using the 21-cm signal observations, the corresponding present-day observed frequency can be expressed as $1420.4/(1+z)$~MHz. The 21-cm differential brightness temperature \cite{Pritchard_2012, Natwariya:2023T}, 
\eq{
\Delta T_b=\frac{T_{\rm exc}-T_{\rm R}}{1+z}\ (1-e^{-\tau_{\nu_0}})\,.\label{eq:dtbe}
}
For the neutral hydrogen gas and the 21-cm line, the excitation temperature ($T_{\rm exc}$) corresponds to the spin temperature ($T_{\rm S}$). In the present study, we do not consider any non-standard radio radiation. Therefore, the background radiation will be solely from CMBR, i.e. $T_{\rm R}=T_{\rm CMB}$. The spin temperature of the neutral hydrogen atoms is defined by the relative number densities in the 1S singlet and 1S triplet hyperfine states. The probability of this transition is once in $\sim 10^7$ years in the absence of any external sources of energy injection. The presence of such sources can significantly affect the transition. In cosmological scenarios, there are three main competitive processes that can affect the evolution of the spin temperature: (I) background radio photons, (II) Ly-$\alpha$ radiation from the luminous sources or exotic sources--- such as decaying/annihilating dark matter, (III) collisions of hydrogen atoms with other hydrogen atoms and residual free electrons or protons,
\eq{
T_{\rm S}^{-1}=\frac{ T_{\rm CMB}^{-1}+x_\alpha\, T_\alpha^{-1}+x_c\, T_{\rm gas}^{-1} }{1+x_\alpha+x_c}\,,
}
where, $T_\alpha$ is the colour temperature of Ly-$\alpha$ radiation due to luminous sources and dark matter annihilation. $T_{\rm gas}$ refers to gas temperature, which indicates the temperature of either protons, electrons, or the neutral particles that remain in thermal equilibrium with each other. $x_\alpha$ is the Ly-$\alpha$ coupling coefficient due to Wouthuysen-Field (WF) effect and $x_c$ is collisional coupling coefficient \cite{Pritchard_2012, Natwariya:2023T, 1952AJ.....57R..31W, Field, 1958PIRE...46..240F, 2006MNRAS.367..259H, 21CMFAST:2011},
\eq{
x_{c} = \frac{T_{\nu_0}}{T_{\rm CMB}}\frac{C_{10} }{A_{10} }\ , \ \  x_{\alpha} = \frac{T_{\nu_0}}{T_{\rm CMB}}\frac{P_{01} }{A_{10} }\ , \label{xcxa}
}
here, $T_{\nu_0}=2\pi\nu_0=68~{\rm mK\,}$ is the corresponding temperature for hyperfine transition. $C_{10}=N_ik_{10}^{iH}$ represents the collision de-excitation rate; here, $i$ stands for hydrogen atom, electron and proton, and $k_{10}^{iH}$ stands for the spin de-excitation specific rate coefficient due to collisions of hydrogen atom with species $i$  \cite{2012RPPh...75h6901P}. $A_{10}=2.85\times 10^{-15}$~sec$^{-1}$ represents the Einstein coefficient for spontaneous emission from triplet to singlet state in neutral hydrogen. $P_{01}=4P_{\alpha}/27$, and $P_\alpha$ is scattering rate of Ly-$\alpha$ radiation \cite{2012RPPh...75h6901P}. The WF coupling coefficient can be written as \cite{2012RPPh...75h6901P, 21CMFAST:2011},
\eq{
x_\alpha=1.7\times10^{11}\,(1+z)^{-1}\, S_\alpha\,J_\alpha\,,
}
here, $S_\alpha=\int dx\, \phi_\alpha(x)\,J_\nu(x)/J_\infty$, is the order of unity and involves atomic physics. It describes the distribution of the photons near Ly-$\alpha$ resonance. $J_\infty$ denotes the flux measured at a distance far from the point of absorption. $J_\alpha$ represents the specific flux of Ly-$\alpha$ photons from astrophysical sources or any other exotic sources of energy injection. Furthermore,  the optical depth of the HI medium can be written as \cite{Natwariya:2023T},
\eq{ 
\tau_{\nu_0} \simeq \frac{3\ n_{\rm HI}}{32\, \pi\, \nu_0^3}\ \frac{T_{\nu_0}}{T_{\rm S}}\ \frac{A_{10}}{H(z)}\ \frac{H(z)/(1+z)}{\partial v/\partial r}\,,
}
here, $n_{\rm HI}\equiv n_{\rm HI}(\bm x,\,z)$ represents the neutral hydrogen number density; where $\bm x$ denotes the position dependence and $z$ denotes the temporal dependence. The number density of neutral hydrogen may change depending on the cell in the simulation Box. $H(z)$ stands for the Hubble parameter. $\partial v/\partial r$ is the proper velocity gradient of neutral hydrogen along the line of sight, and it can be taken as $H(z)/(1+z)$ for high redshift or in the absence of peculiar velocity. Furthermore, the term $\big([H(z)/(1+z)]/[\partial v/\partial r]\big)$ in the above equation can be written as $\big(H(z)/[\partial v_r/\partial r+H(z)]\big)$ in the comoving coordinates; where $\partial v_r/\partial r$ is the comoving gradient of comoving velocity of neutral hydrogen along the line of sight component \cite{Mesinger:2011FS}. In the {\tt 21cmFAST} code, $\partial v_r/\partial r$ is calculated by first-order density perturbations using the Zel`Dovich approximation. Therefore, this quantity also has spatial dependence. The optical depth of the hydrogen medium for the 21-cm line is very small, $\tau_{\nu_0}\ll1$. In this limit, the 21-cm differential brightness temperature (eq. \ref{eq:dtbe}) can be written as,
\eq{
\Delta T_b\simeq  &27\,x_{\rm HI}\,(1+\delta_b)\left(1-\frac{T_{\rm CMB}}{T_{\rm S}}\right) \left(\frac{\Omega_{\rm b}\,h^2}{0.023}\right) \nonumber\\
&\times \sqrt{\frac{0.15}{\Omega_{\rm m }\,h^2}\ \frac{1+z}{10}}\ \left(\frac{H(z)/(1+z)}{\partial v/\partial r}\right) ~{\rm mK},
\label{eq:DeltaTb}
}
here, $x_{\rm HI}$ and $\delta_b$ are neutral hydrogen fraction and baryon energy density contrast, respectively. The 21-cm differential brightness temperature is commonly referred to as the 21-cm signal; for clarity, we will use the term 21-cm signal from now on.

The dimensionless 21-cm signal power spectrum can be defined as,
\eq{
\Delta_{21}^2(k,\,z)= \frac{k^3}{2\,\pi^2\, V}\ \langle\Delta T_b(\bm k,\,z)\Delta T_b^*(\bm k,\,z)\rangle_k\,,\label{eq:powersp}
}
here, $V$ stands for the volume of the box and $\Delta T_b(\bm k,\,z)$ represents the Fourier transformation of $\Delta T_b(\bm x,\,z)$. This equation is also equivalent to $\Delta_{21}^2(k,\,z)= {k^3}/({2\,\pi^2\, V})\times \bar{\Delta T_b}(z)^2 \langle|\delta_{21}(\bm k,\,z)|^2\rangle_k$ where $\delta_{21}(\bm x,\,z)=\Delta T_b(\bm x,\,z)/\Bar{\Delta T_b}(z)-1$, and $\bar{\Delta T_b}(z)$ is the the average 21-cm signal of the box at corresponding redshift or the global 21-cm signal. To get the $\bar{\Delta T_b}(z)$ of the box at redshift $z$, we can add the $\Delta T_b(\bm x,\,z)$ from every cell in the box and then divide by the total number of cells at that redshift. In the next subsection, we discuss how dark matter annihilation injects energy into the surrounding medium and affects the evolution of the 21-cm signal.

\subsection{Energy injection and deposition into the medium by dark matter annihilation}
\label{ssec:energyinjection}
Dark matter annihilation can inject energy into the plasma via energetic particles. In the present study, we focus on two channels: (I) electron/positron and (II) photons. Dark matter can decay into various Standard Model particles. However, for the relevant timescale, the resulting products will mainly include electrons/positrons, photons, and neutrinos. The effect of neutrinos on the thermal and ionization evolution of plasma can be neglected due to weak interactions. The energy injection per unit time and per unit volume,
\begin{alignat}{2}
	\frac{dE^{\rm inj}}{dV\,dt} = \,\underbrace{\bar\rho_{\rm0,\DM}^2\ (1+z)^6\,(1+\delta(\bm x,\,z))^2}_{\text{physical DM energy density}}\ f_{\rm DM}^2\,\frac{\langle \sigma\,v\rangle}{m_{\rm DM}}\,,\label{eq:inj}
\end{alignat}
here,  $\bar\rho_{\rm 0,\DM}$ stands for the present-day dark matter energy density, and $\delta(\bm x,\,z)$ stands for the density contrast. $f_{\DM}$ is the dark matter fraction that annihilates. Since we consider that all dark matter consists of identical particles or is a single component, $f_{\DM}$ will be 1. Furthermore, $\sv$ counts for the thermal average of the cross-section and relative velocity of dark matter particles, and $m_{\DM}$ represents the mass of the dark matter. In the equation, the first three factors represent the physical energy density of the dark matter at the time of energy injection. The density contrast, $\delta(\bm x,\,z)$, can be obtained from each cell in the box. In a single cell, the energy injection will be $\propto (1+\delta_{\rm cell})^2$, where $\delta_{\rm cell}$ is the dark matter density contrast of that cell. This can be written as $(1+2\delta_{\rm cell}+\delta_{\rm cell}^2)$. For the total energy injection at a redshift $z$, we need to sum it over all cells: $\sum_{\rm cell}(1+2\delta_{\rm cell}+\delta_{\rm cell}^2)$. The term $\sum_{\rm cell}\delta_{\rm cell}$ represents the sum of density contrasts in all the cells. By definition, $\delta_{\rm cell}$ is the fluctuation around the mean density, so its average over all cells in a large volume will be zero. Therefore, the total energy injection at a given redshift will be $\propto N_{\rm cell}(1+\langle\delta^2({\bm x})\rangle)$, where $\langle\delta^2({\bm x})\rangle$ represents the variance of the density contrast: $1/N_{\rm cell}\times \sum_{\rm cell}\delta^2_{\rm cell}$. This is also referred to as the dark matter annihilation boost factor, which arises from the formation of structures in the Universe. For a homogeneous energy injection, this boost factor is computed once at a fixed redshift and is applied uniformly across space. Analytical models for calculating the boost factor use some kind of density profile, such as the Navarro-Frenk-White (NFW) density profile and the Einasto profile. However, in the case of inhomogeneous energy injection, it is necessary to monitor the density contrast of each individual cell. 

The simplest process for energy deposition involves ``on-the-spot" approximation in which energetic particles deposit their energy into the plasma before they can travel to a considerable length scale. In this scenario, the rate of injection and deposition events per unit volume by dark matter annihilation becomes the same and is given by \cite{Sun:2023},
\eq{
	\frac{dN^{\rm inj/dep}}{dV\,dt} = \bar\rho_{\rm0,\DM}^2\ (1+z)^6\,(1+\delta(\bm x,\,z))^2\,\frac{\sv}{m_{\rm DM}^2}\,.\label{eq:eventinjdep}
}
In this case, the energy deposition by dark matter annihilation does not depend on the annihilation channel, whether it annihilates into electrons/positrons or photons. However, in reality, the surrounding plasma may not absorb all of the injected particles efficiently, and the deposition efficiency can, for instance, depend on the annihilation products, their energy spectrum, and the ionisation state of the medium. 

The energy loss by electrons/positrons is dominated by the inverse-Compton (IC) scattering process at energies above 1~MeV. Below this energy, the electrons and positrons lose their energy mainly by ionizations and excitations of HI, HeI, and HeII. For instance, the mean free path for inverse-Compton scattering is approximately $\mathcal{O}(10^{-6})$ Mpc at redshift $z\sim10$ \cite{Evoli:2012}. For an electron energy of approximately $10^6$~eV, the mean free path for ionization and excitation is roughly $\mathcal{O}(10^{-2})$~Mpc at redshift $z\sim10$. Moreover, a magnetic field with a strength as low as $10^{-20}$ G can confine electrons/positrons within a radius of less than approximately 0.05~Mpc \cite{Sun:2023}, and the intergalactic magnetic fields are anticipated to have strengths larger than this, typically order of \(\sim10^{-16}~\rm G\) \cite{Fermi_LAT:2018AB, Tavecchio:2010GFB}. We take the cell size to be 0.75~Mpc, and therefore, energy deposition by even ultrarelativistic electrons can be treated in an on-the-spot manner. These high-energy electrons can also upscatter the cosmic microwave background (CMB) photons, 
known as secondary photons, and can be included in the outgoing spectrum of photons depending on their energy. On the other hand, positron annihilation can also produce secondary photons. Consequently, the energy deposition by electrons/positrons can be characterized mainly by two transfer functions: I) the energy they contribute to heating, ionization, and Ly$\alpha$ excitation over a redshift step $\Delta z$,
\eq{ 
\begin{bmatrix}
\Delta T_{\rm gas} \\
\Delta x_e \\
J_\alpha
\end{bmatrix}_{\rm DM}
&= D_{ce}(\delta, x_{\text{HI}}|\Delta z) \frac{dN_{e}^{\text{in}}}{dE}\,,\label{dTe}}
and II) the outgoing spectrum of photons (for more details, please refer to the article \cite{Sun:2023}),
\eq{
\frac{dN_{\gamma}^{\text{out}}}{dE} &= T_{\gamma e}(\delta, x_{\text{HI}}|\Delta z) \frac{dN_{e}^{\text{in}}}{dE}\,,\label{eq:getrans}
}
here, $D_{ce}$ maps an input spectrum of electrons/positrons from dark matter annihilation into the energy deposition into different channels (heating of the gas in the cell: $\Delta T_{\rm gas}$, ionization of the gas: $\Delta x_e$, and Ly$\alpha$ excitation of neutral Hydrogen: $J_\alpha$). The input spectrum, $dN_i^{\rm in}/dE$, is defined in terms of the number density of emitted particles per unit energy and average baryon density,
\eq{\frac{dN_i^{\rm in}}{dE}=\frac{1}{\bar n_b}\times \frac{dN_i^{\rm in}}{dE dV}\,,}
here, $i$ index stands for the photons, $\gamma$, and electrons/positrons, $e$. Further, $T_{\gamma e}$ in equation \eqref{eq:getrans} maps an input electron/positron spectrum into an outgoing photon spectrum. These transfer functions are evaluated for each cell based on baryonic overdensity, which is determined through the local density contrast, and the total local ionization fraction, which is 1-$x_{\rm HI}$. Where $x_{\rm HI}$ represents the neutral hydrogen fraction of the cell. At high energies, electrons and positrons behave similarly, having identical IC scattering cross-sections. Therefore, the term ``electrons” refers to both electrons and positrons throughout this paper. The outgoing upscattered photons due to the IC process can have different energies depending on the energy of the electrons. We stay well within the Thomson limit during the relevant range of redshift. This means that the energy of the incoming photons, here CMB photons, is much smaller than the rest mass of the electrons/positrons in their rest frame \cite{Valds:2010}. This sets the upper and lower limits on the energy of upscattered CMB photons, ranging from \(E_{\rm in}^\gamma\) to \(4\gamma_e^2E_{\rm in}^\gamma\). Here, \(E_{\rm in}^\gamma\) represents the energy of the initial photon, while \(\gamma_e\) denotes the Lorentz factor of the electron. For example, the CMB photons have an average energy of approximately \(7 \times 10^{-3}\) eV at a redshift of 10. In the rest frame of an electron with a total energy of 1 TeV, the maximum energy of the incoming photon will be about \(\gamma_e E_{\text{in}}^\gamma \sim 10^4\) eV. This value is much smaller than the rest mass of the electron, ensuring the Thomson limit. Therefore, at a redshift of 10, an electron having the energy of 1~MeV can upscatter CMB photons in the energy range from $\sim0.007$~eV to $\sim0.1$~eV, which is insufficient for the ionization and excitation. Further, an electron with an energy of 100~MeV can upscatter CMB photons up to an energy of $\sim1$~KeV, which is sufficient to ionize the atoms. In contrast, electrons having the energy of 1~GeV can upscatter CMB photons to an energy of $>100$~KeV, which can free stream into the plasma until their energy is redshifted sufficiently to be absorbed in the plasma.

Similar to electrons, photons also deposit their energy into plasma through different processes. At higher energies above $\sim10^8$~eV, the larger fraction of energy deposition happens through pair production. In the energy range of photons from $\sim10^4$~eV to $\sim10^8$~eV, energy deposition is dominated by the Compton interaction. At energies below $\sim10^4$~eV, photoionization interaction dominates over other processes \cite{Evoli:2012}. However, for energies exceeding \(10^3\) eV at redshift \(z\sim 10\), the mean free path for these processes remains much larger than the Hubble radius. It is only below \(10^3\) eV that the mean free path for photoionization becomes smaller than the Hubble radius at \(z \sim 10\). Similarly, at \(z \sim 100\), the mean free path for Compton scattering also falls below the Hubble radius for energies below approximately \(10^6\) eV. Nevertheless, for the range of dark matter masses considered, the mean free path of emitted photons remains significantly larger than the Hubble radius at the redshift range of interest. Therefore, the transfer function can be classified into three categories: I) $D_{c\gamma}$--- mapping an input spectrum of photons per baryon to energy deposition into the heating, ionization, and Ly$\alpha$ excitation over a redshift step $\Delta z$,
\eq{ 
\begin{bmatrix}
\Delta T_{\rm gas} \\
\Delta x_e \\
J_\alpha
\end{bmatrix}_{\rm DM}
&= D_{c\gamma}(\delta, x_{\text{HI}}|\Delta z) \frac{dN_{\gamma}^{\text{in}}}{dE}\,,\label{dTg}}
II) \(P_{\gamma\gamma}\)--- propagating photon transfer function that maps the input spectrum of photons which do not go under any scattering process over a redshift interval of $\Delta z$, and III) \(T_{\gamma\gamma}\)--- maps an input spectrum of photons to the spectrum of outgoing photons from scattering events \cite{Sun:2023},
\eq{
\frac{dN_{\gamma}^{\text{out}}}{dE} &= \left[P_{\gamma \gamma}(\delta, x_{\text{HI}}|\Delta z) + T_{\gamma \gamma}(\delta, x_{\text{HI}}|\Delta z)\right]\,\frac{dN_{\gamma}^{\text{in}}}{dE}\,.\label{eq:ggtrans}
}
If the mean free path of the photons at the time of emission is small compared to the size of the cell, they maintain their inhomogeneities--- as they can deposit their energy promptly. It includes photons with energies ranging from about 10.2~eV to $\sim100$~eV. Photons with energies ranging from about 100 eV to 10 keV can have mean free paths larger than the cell size, allowing them to travel to neighbouring cells. The spectrum incident on each cell varies depending on the spatial distribution of dark matter in the box and must be calculated by integrating along the light cone, taking into account redshifting and the energy absorption by the medium. At higher energies, the cooling distance of photons becomes larger than the Hubble length scale, extending beyond the size of the simulation box. These photons are treated as a homogeneous injection spectrum for subsequent redshifted boxes. However, the energy deposition by these photons in the subsequent redshifted boxes depends on the density of the baryons in the corresponding cells. For more information on the treatment of transfer functions of electrons/positrons and photons, please refer to the article \cite{Sun:2023}.

\subsection{Thermal and Ionization evolution of the gas}\label{subsec:tgas}

We assume a homogeneous evolution above a redshift of 45, since dark matter annihilation has a relatively small impact on local variations in the thermal and ionization history during this epoch due to the absence of luminous structure formation. 
In the presence of energy injection by dark matter annihilation at \(z>45\), the thermal evolution of the baryons with redshift can be written as \cite{1999ApJ...523L...1S, 2000ApJS..128..407S, Cheung:2018vww, Xu:2024uas, Zhao:2025ddy, 2019JHEP...02..187C, 2018PhRvL.121a1103D, 2009PhRvD..80b3505G}, 
\eq{
    \frac{dT_{\rm gas}}{dz}  =  \frac{2\,T_{\rm gas}}{(1+z)} + \frac{\Gamma_{C}}{(1+z)\,H}\, (T_{\rm gas}-T_{\rm CMB})+\frac{dT_{\rm gas}}{dz}\bigg|_{\DM}\,.
	\label{dtdz} 
}
The Compton scattering rate in the above equation is defined as,
\begin{equation}
	\Gamma_{C}= \frac{8\, \sigma_T\, a_r T_{\rm CMB}^4\, x_e}{3\,(1+f_{\rm He}+x_e)\,m_e}\,,
\end{equation}
where, $\sigma_T$, $m_e$, and $a_r$ represent the cross-section for Thomson scattering, mass of the electron, and Stefan-Boltzmann radiation constant, respectively. Furthermore, $x_e=n_e/n_{\rm H}$ denotes the ionization fraction, where $n_e$ and $n_{\rm H}$ are the free electron and total hydrogen number densities, respectively. The last term represents the contribution from dark matter annihilation,
\eq{
 \frac{dT_{\rm gas}}{dz}\bigg|_{\DM} = - \frac{1}{(1+z)\,H}\ \frac{2}{3\,n_{\rm H}\,(1+f_{\rm He}+x_e)}\ \frac{dE^{\rm dep,heat}}{dV\,dt}\,,
}
here, $f_{\rm He}=n_{\rm He}/n_{\rm H}$ stands for the Helium fraction, and \(n_{\rm He}\) represents the number density of helium. The energy deposition by dark matter annihilation \cite{Liu:2016}, 
\begin{alignat}{2}
    \frac{dE^{{\rm dep},c}}{dV\,dt} = f_c(z,\,m_{\DM})\ \frac{dE^{\rm inj}}{dV\,dt}\,, \label{eq:dep}
\end{alignat}
here, function $f_c(z,\,m_{\DM})$ denotes the efficiency of energy deposition into the plasma via a specific channel, $c$--- ``heat" (heating), ``ion" (ionization), and ``exc" (excitation). This function also varies for different species (electrons/positrons and photons) injected by dark matter annihilation \cite{Liu:2016}. In the presence of exotic energy injection into the plasma, evolution of ionization fraction can be written as  \cite{1999ApJ...523L...1S, 2000ApJS..128..407S, 2018PhRvL.121a1103D, 2011PhRvD..83d3513A, 2009PhRvD..80b3505G},
\eq{
\frac{dx_e}{dz} &= \frac{\mathcal{P}}{H\,(1+z)}\times\,
\Big[ \,n_{\rm H}\, x_e\, x_{\rm HII}\,\alpha_{\rm H} \nonumber \\
&\hspace{1.5em}- 4\,(1-x_{\rm HII})\,\beta_{\rm H} \,e^{-E_{\alpha}/T_{\rm CMB}} \Big]+\frac{dx_e}{dz}\bigg|_{\rm DM}\,,\label{dxedz}
}
here, $\mathcal{P}$ represents the Peebles coefficient which indicates the decay probability of a hydrogen atom from the $n=2$ state to the ground state before photoionization occurs \cite{Peebles:1968ja, AliHaimoud:2010dx, 2018PhRvL.121a1103D},
\begin{alignat}{2}
    \mathcal{P}=\frac{1+K_{\rm H}\,\Lambda_{\rm H}\,n_{\rm H}\,(1-x_{\rm HII})}{1+K_{\rm H}\,(\Lambda_{\rm H}+\beta_{\rm H})\,n_{\rm H}\,(1-x_{\rm HII})\,}\,,
	\label{}
\end{alignat}
here, $K_{\rm H}=\pi^2/(E_\alpha^3\, H)$ account for redshifting of Ly-$\alpha$ photons due to expansion of the Universe, and $\Lambda_{\rm H}=8.22/{\rm sec}$ represents the 2S-1S level two photon decay rate of hydrogen atom \cite{Tung:1984}. Further, $x_{\rm HII}$, $\alpha_{\rm H}$ and $\beta_{\rm H}$ denote the free proton fraction, case-B recombination coefficient and photoionization rate, respectively \cite{1999ApJ...523L...1S, 2000ApJS..128..407S}. Free proton fraction can be taken as $x_{\rm HII}\equiv x_e$ in the absence of helium.  $E_\alpha=(3/4)\, E_0$ stands for the Ly-$\alpha$ transition energy for the hydrogen atom, and $E_0=13.6$~eV is the ground state binding energy of the neutral hydrogen atom. The contribution in the ionization fraction of the gas due to the dark matter annihilation,
\eq{
\frac{dx_e}{dz}\bigg|_{\rm DM} &=  - \frac{1}{H\,(1+z)} \ \frac{1}{n_{\rm H}}\ \Bigg[\,\frac{1}{E_0}\, \frac{dE^{\rm dep,ion}}{dV\,dt} \nonumber \\
        &\hspace{8.5em}+\frac{1-\mathcal{P}}{E_\alpha}\, \frac{dE^{\rm dep,exc}}{dV\,dt}\,\Bigg]\,.
}

For the dark matter annihilation scenario, there will be three main contributions for Ly$\alpha$ flux: (i) excitation of neutral hydrogen by stellar X-ray, (ii) stellar emission occurring between  Ly-$\alpha$ and Lyman limit, (iii) Ly-$\alpha$ flux from collisional excitation/deexcitations of neutral hydrogen, caused by energy injection into the plasma due to dark matter annihilation. In the presence of exotic energy injection, Ly-$\alpha$ flux can be written as,
\begin{alignat}{2}
J_{\alpha,\,\rm DM}= \frac{1}{8\pi^2\nu_\alpha^2\,H^2}\ \frac{dE^{\rm dep,exc}}{dV\,dt}\,,
\end{alignat} 
here, $\nu_\alpha$ stands for Ly-$\alpha$ frequency. 

The solutions of the above thermal and ionization equations (\ref{dtdz} and \ref{dxedz}), work as initial conditions ($T_{\rm gas}$ and $x_e$ at $z=45$) for {\tt 21cmFAST} code linked with {\tt DM21cm}. These coupled differential equations with the presence of exotic energy injection from dark matter are solved using {\tt DarkHistory} code. After a redshift of about 45, the spatial distribution of baryons significantly influences the local evolution of the gas. Therefore, in order to obtain the kinetic temperature and ionization fraction of the gas in the cells of the box, it is necessary to keep track of the local state of the gas. Following the structure of {\tt 21cmFAST}, the thermal and ionization  equations, (\ref{dtdz} \& \ref {dxedz}), are modified as  \cite{Mesinger:2011FS, Sun:2023},
\eq{
\frac{dT_{\rm gas}(\bm x,\,z)}{dz} &= \frac{2\,T_{\rm gas}}{3\,n_A}\ \frac{dn_A}{dz}-\frac{T_{\rm gas}}{1+x_e}\ \frac{dx_e}{dz} \nonumber\\ 
&-\frac{1}{(1+z)\,H} \ \frac{2}{3\,(1+x_e)}\sum_p \epsilon_p +\frac{dT^{\rm DM}_{\rm gas}}{dz}\,,
}
\eq{
\frac{dx_e(\bm x,\,z)}{dz} &= \frac{1}{(1+z)\,H}\ \big[\alpha_ACx_e^2n_Af_{\rm H}-\Lambda_{\rm ion} \big] + \frac{dx^{\DM}_e}{dz}\,,
}
here, $n_A\equiv n_A(\bm x,\,z)$ is the baryon number density representing total number density of hydrogen and helium nuclei; $\epsilon_p\equiv \epsilon_p(\bm x,\,z)$ stands for heating/cooling rate per nucleus including the Compton and X-ray heating of the gas; $\alpha_A$ denotes the case-A recombination coefficient; $C=\langle n_e^2\rangle/\langle n_e\rangle^2$ denotes the clumping factor for the free-electrons on the scale of the simulation cell; $f_{\rm H}=n_{\rm H}/(n_{\rm H}+n_{\rm He})\,$ is the fraction of the hydrogen nucleus; and $\Lambda_{\rm ion}$ represents the ionization rate per nucleus accounting the radiation which is emitted at \((\bm x',\, z')\) and propagates to \(\bm x\) before ionizing or getting absorbed into the plasma. Energy deposition into cells in form of kinetic temperature (\({dT^{\rm DM}_{\rm gas}}/{dz}\)) and ionization (\({dx^{\DM}_e}/{dz}\)) by dark matter over a redshift interval of \(\Delta z\) is calculated using the equations (\ref{dTe} and \ref{dTg}) depending on the annihilation channel. The contribution in the Ly$\alpha$ flux due to dark matter annihilation is also given in the equations (\ref{dTe} and \ref{dTg}). 

{It is to be noted that the soft photon heating, arising from free-free transitions in the presence of a soft photon background, is neglected here for simplicity \cite{Cyr:2024vkt}. As this term plays a strongly subdominant role in the evolution of the gas temperature in the absence of any external soft photon background, where the CMB is the sole source of low-frequency photons. Another potential source of soft photons could be the first astrophysical sources, which would represent a highly local process. The present work focuses on exploring the effects of homogeneous and inhomogeneous dark matter annihilation on the 21-cm signal; the inclusion of free-free heating would modify the local evolution of the gas temperature and background radiation equally in both scenarios. However, in future work, it would be interesting to explore how this term modifies the 21-cm signal for both homogeneous and inhomogeneous dark matter annihilation scenarios. In the context of the present work, a third source of soft photons could be redshifted or secondary photons from dark matter annihilation. However, we consider dark matter with a mass of 1 MeV or larger, such that the redshifted photons can have an energy of about 1 keV at $z = 0$ when annihilation occurs at $z \sim 1000$, which is much larger than the cut-off scale of 0.235 eV for soft photons at $z \sim 0$. Notably, this type of contribution would have homogeneous effects rather than inhomogeneous effects. Similarly, the secondary photons also have a mean free path larger than the box size, contributing homogeneously as discussed earlier.}

\section{Machine Learning Method}
\label{sec:ML}
This section provides a comprehensive description of the image classification framework, which utilizes a Convolutional Neural Network architecture for discriminating between different models. We present the network architecture design in detail, including its layer configurations and key computational components. Furthermore, we outline the training methodology employed to optimize the model parameters and performance.

\subsection{CNN model}
\label{ssec:CNN}
This paper employs a standard and simple convolutional neural network (CNN) architecture, commonly used in prior work. The input data may consist of either 2D pixels (defined on an $N\times N$ grid) or 3D voxels (defined on an $N\times N\times N$ grid), both starting as single-channel inputs. The CNN comprises four convolutional layers, each following the same sequence of operations: a $3\times3$ convolution, Leaky Rectified Linear Unit (ReLU) activation, $2\times2$ max pooling with a stride of 2, and batch normalization.

The first convolutional layer transforms the single input channel into 32 feature channels. This is followed by subsequent layers that progressively increase the channel depth to 64, 128, and finally 256, enhancing the ability of the network to capture hierarchical features. In each layer, we adopt the activation by Leaky ReLU, max pooling with a kernel size and stride of 2, and batch normalization. Before transitioning to the fully connected layer, a dropout layer is applied with a 50 percent probability to prevent overfitting by randomly deactivating half of the neurons during training. The final output is a single scalar value constrained between 0 and 1, generated by a sigmoid activation function. This output serves as a binary classifier: a prediction of 1 indicates that the input corresponds to the inhomogeneous energy injection/deposition, while a prediction of 0 classifies it as the homogeneous energy injection model. The model's output is a probability between 0 and 1, indicating the likelihood of the image belonging to a given class. We apply a threshold to classify the image as either a homogeneous or an inhomogeneous energy injection model.

In this study, we investigate various CNN architectures designed to effectively extract meaningful information from input images while minimizing the loss of information content. The most basic configuration involves using raw image batches as the sole input. Our analysis reveals that the 21-cm brightness temperature maps generated from inhomogeneous and homogeneous annihilation scenarios exhibit visual similarities. Upon closer inspection, the fluctuation patterns between the two models show minimal discernible differences, with the most notable distinctions lying in their global amplitude and scatter.

This observation implies that when the images are normalized—by subtracting the mean and dividing by the standard deviation—the resulting representations from both models become nearly indistinguishable. Since such normalization is inherently performed during batch normalization in CNNs, critical distinguishing features may be suppressed. To address this, we introduce an additional preprocessing step: We explicitly extract the mean and standard deviation of the original images and concatenate these statistical measures with the latent feature vector before passing them to the fully connected layers.
We find that when the training dataset is limited in size, incorporating these supplementary statistics provides a marginal but measurable improvement in the model's ability to discriminate between the two scenarios. {The primary job of a model is to learn the underlying distribution of the data. The mean and standard deviation are summary statistics of that distribution. With a very small dataset, the model has a poor estimate of the true population mean and standard deviation. Explicitly including these additional statistics provides them with a ``hint" or a ``head start." However, as the training set grows larger and more representative of the true data-generating process, the model can directly and robustly estimate the mean and standard deviation from the raw data itself. The explicit features become redundant because the model has already internalized that information.

A model neural network has a limited capacity to learn. With a small dataset, there is ``free capacity." Using some of that capacity to incorporate and learn from the mean and standard deviation features is a good use of resources. As the dataset grows, the model must use its finite capacity to learn more subtle, complex, and higher-order patterns from the raw data. The simple summary statistics become relatively less important and can even be a distraction if they crowd out the learning of more discriminative, non-linear features.}


\subsection{Fine Tuning and Training}
\label{ssec:training}
The most challenging aspect of machine learning methods is tailoring the architecture to the specific problem and dataset. Hyperparameters can be optimized systematically using techniques like grid search, Latin Hypercube sampling, Bayesian optimization, or genetic algorithms \cite{GA1992}. In this study, we empirically determine a hyperparameter combination rather than exhaustively seeking the optimal configuration. While not necessarily ideal, this set performs adequately for our simulations and can be further refined for practical use.

We prepare 1,000 realizations for the conventional model with homogeneous energy injection/deposition, and also for inhomogeneous energy injection/deposition for every dark matter annihilation model. Where each realization has a box of 48 Mpc on a side, with a voxel resolution of 0.75 Mpc. {This specific choice of box size ($L_\mathrm{box}$) and number of grids in a box ($N_\mathrm{grid}$) was the result of a systematic exploration to optimize for both accuracy and computational feasibility. We tested a range of configurations: 0.75 Mpc ($N_\mathrm{grid} = 64^3$, $L_\mathrm{box} = 48$ Mpc), 1.0 Mpc ($N_\mathrm{grid} = 64^3$, $L_\mathrm{box} = 64^3$ Mpc), 1.5 Mpc ($N_\mathrm{grid} = 64^3$, $L_\mathrm{box} = 96$ Mpc), 1.5 Mpc ($N_\mathrm{grid} = 200^3$, $L_\mathrm{box} = 300$ Mpc), 1.75 Mpc ($N_\mathrm{grid} = 64^3$, $L_\mathrm{box} = 112$ Mpc), 2.0 Mpc ($N_\mathrm{grid} = 128^3$, $L_\mathrm{box} = 256$ Mpc), and 5.0 Mpc ($N_\mathrm{grid} = 64^3$, $L_\mathrm{box} = 320$ Mpc). After comparing the results, we concluded that the configuration with $N_\mathrm{grid} = 64^3$ and $L_\mathrm{box} = 48$ Mpc (voxel resolution of 0.75 Mpc) provided the optimal balance, delivering sufficiently detailed maps while remaining computationally feasible for the large number of simulations required in this work.} The size of the voxel corresponds to the angular size of $3.5$ arcmin at redshift $z=12$ (it corresponds to the line of sight resolution, in terms of a frequency resolution, of 150 kHz at 1.4 GHz). The 48 Mpc simulated box is divided into $4\times 4\times 4$ subregions, where each 3-D image has $16\times 16\times 16$ voxels. In total, we have 128,000 images: half of them are for the homogeneous energy injection model, and the rest are for the inhomogeneous model.
We randomly select 20\% of the data as a hold-out test set, and use the remaining 80\% as the reference sample. This reference sample is further divided into a training set (80\%) and a validation set (20\%). The training set is used to optimise the model, while the validation set is used to assess model performance and monitor overfitting. Notably, the validation set is not involved in the optimisation process. Training is stopped when the validation loss ceases to improve, indicating the potential onset of overfitting.


\begin{figure*}[]
    \centering
    \includegraphics[width=\textwidth]{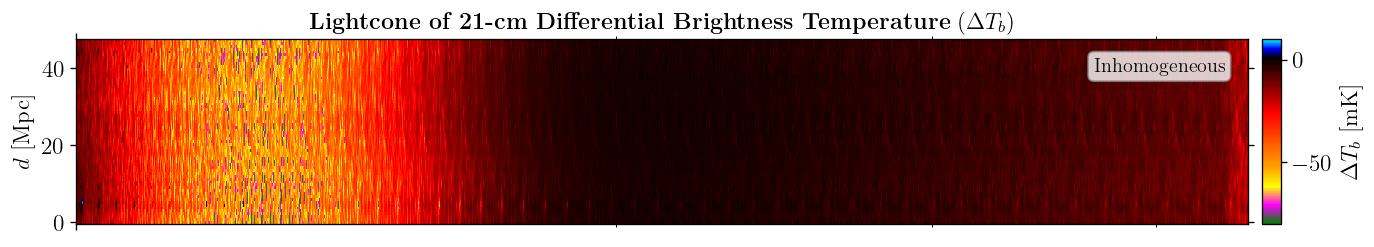}
    \includegraphics[width=\textwidth]{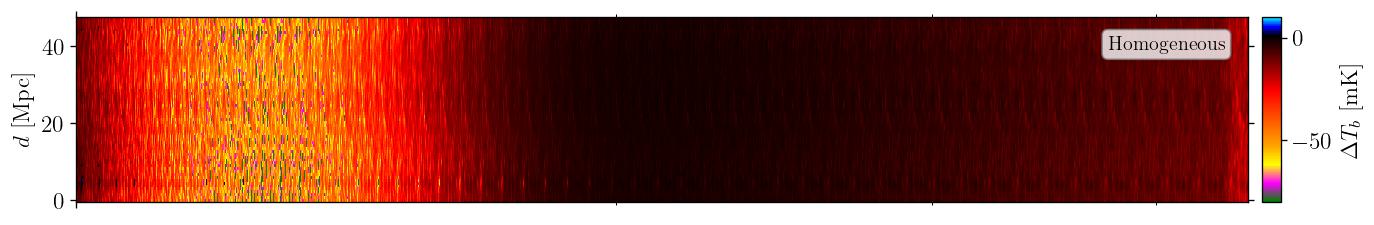}
    \includegraphics[width=\textwidth]{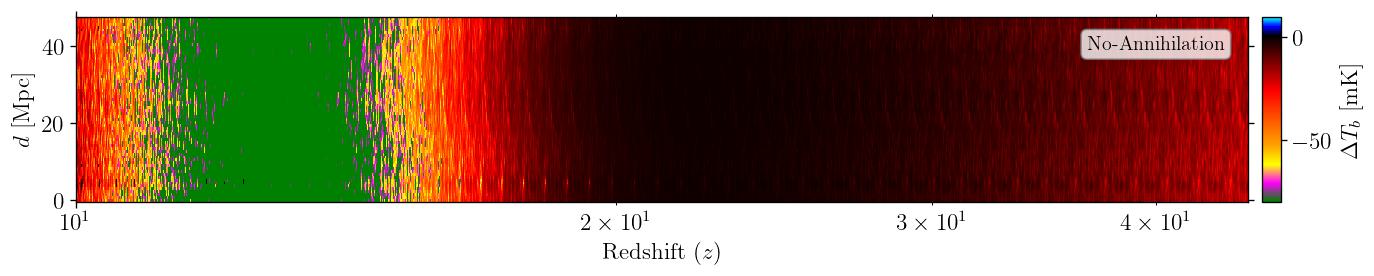}   
    \caption{21-cm differential brightness temperature lightcone from redshift 45 to 10 for dark matter annihilation and no-annihilation scenario. The lower panel represents the conventional $\Lambda$CDM model without dark matter annihilation. The middle and the top panel represent the annihilation scenarios with the dark matter particle mass \(m_{\DM} = 1\)~MeV, and cross section $\sv=1\times10^{-29}~{\cms}$ for the electron/positron channel. The upper panel illustrates inhomogeneous energy injection and deposition, while the middle panel depicts the scenario with homogeneous energy injection and deposition.}
    \label{fig:lightcone}
\end{figure*}

\section{Results and Discussion}
\label{sec:result}
We start by examining dark matter annihilation through the electron-positron channel ($\ee$) channel with $m_\DM = 1$ MeV and cross-section $\sv = 1\times 10^{-29}~\cms$. This mass is selected for its efficiency in on-the-spot energy absorption by the gas for our redshift range of interest. The cross-section $\sv = 10^{-29}~\cms$ is motivated by current observational upper limits for $m_\DM = 1$~MeV from X-ray and CMB data \cite{Cirelli:2023tnx, 2016PhRvD..93b3527S}, while remaining sufficiently large to produce visually discernible signatures in the 21-cm maps for demonstration. We, therefore, adopt this ($m_\DM = 1$ MeV, $\sv = 10^{-29}~\cms$, $\ee$) scenario as our fiducial model. We also discuss other parameter sets in the subsequent sections.

\begin{figure*}
    \begin{center}
        \subfloat[] {\includegraphics[width=2.1in,height=2.1in]{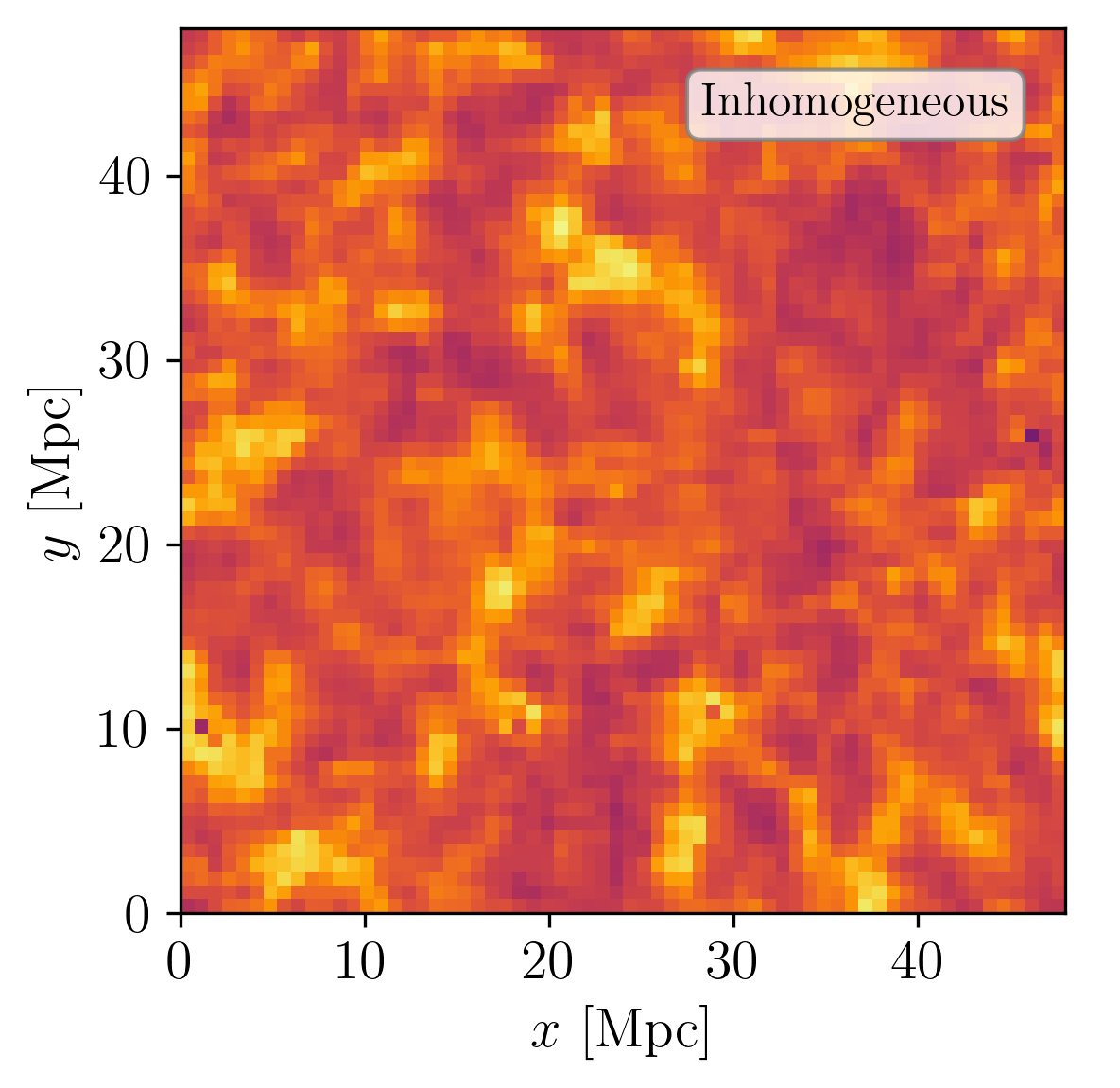}\label{fig:slice_YInHom}}
        \subfloat[] {\includegraphics[width=2.3625in,height=2.1in]{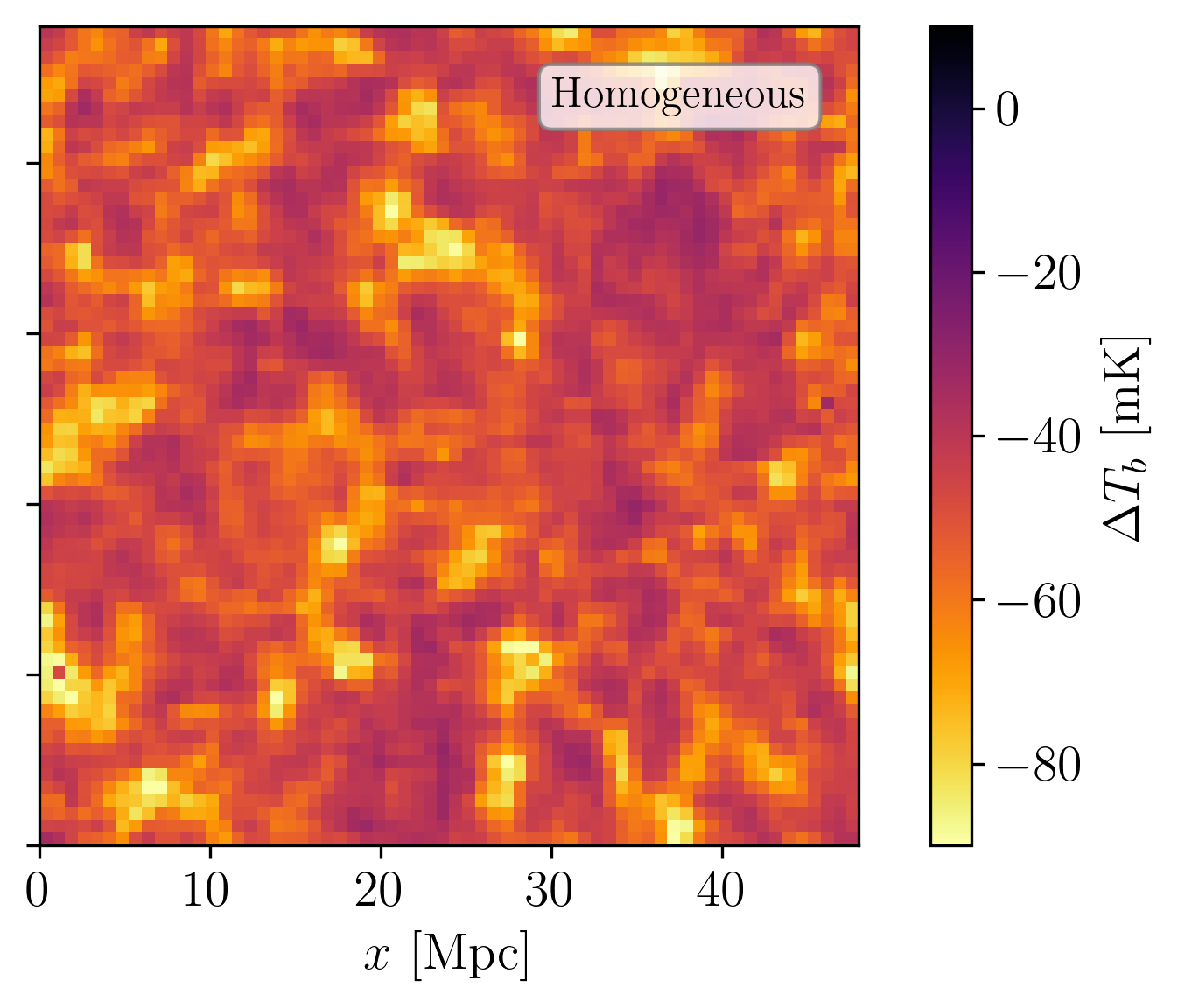}\label{fig:slice_NInHom}} 
        \subfloat[] {\includegraphics[width=2.625in,height=2.15in]{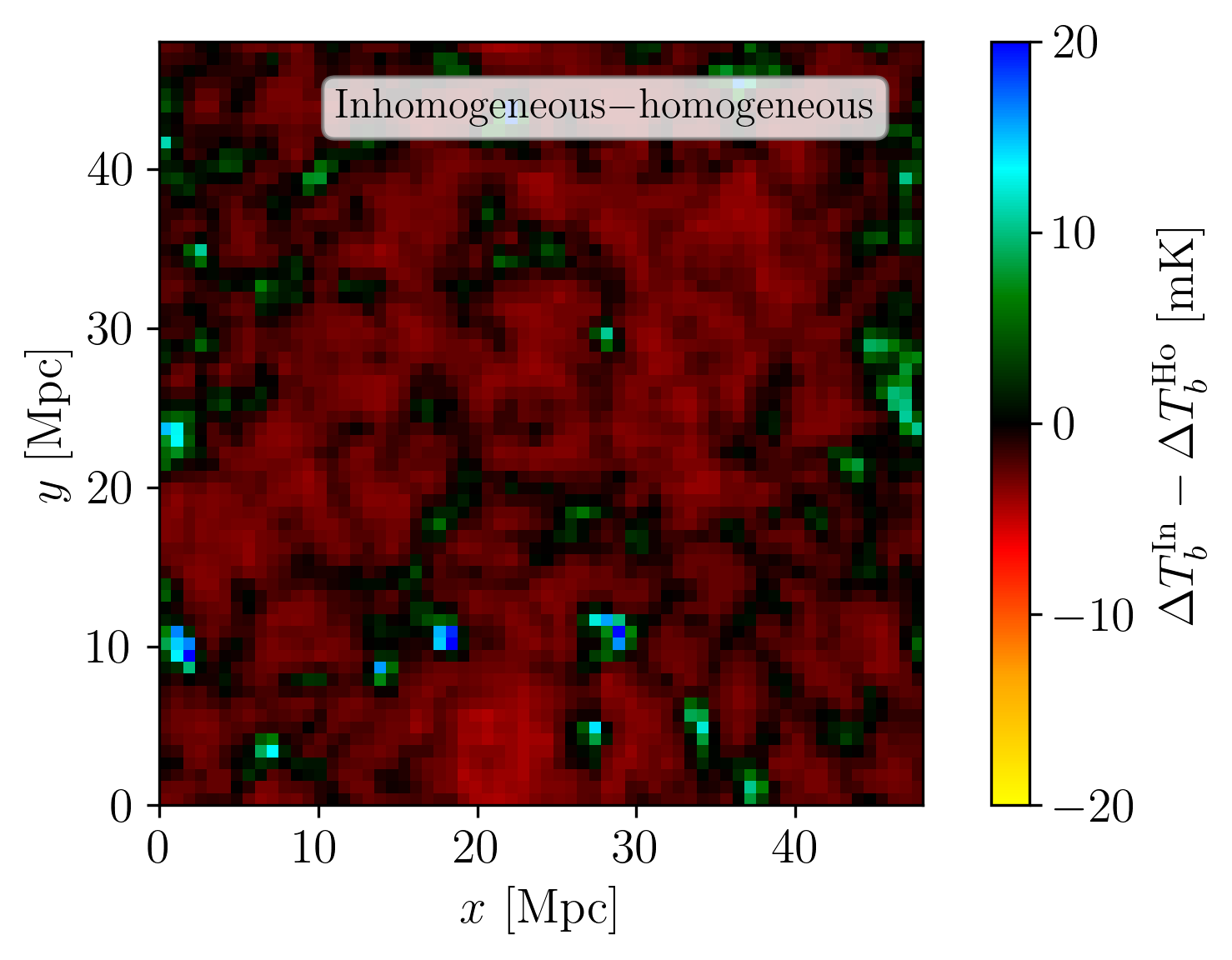}\label{fig:slice_in_ho}}
    \end{center}
    \caption{21-cm differential brightness temperature snaps at redshifts 12 for dark matter particle mass $m_{\rm DM}=1$~MeV, and cross-section $\sv=1\times10^{-29}~\cms$ for electron/positron channel. Left: Inhomogeneous energy injection/deposition into the medium. Middle: homogeneous energy injection/deposition into medium. Right: Difference between 21-cm differential brightness temperature snaps of inhomogeneous and homogeneous energy injection/deposition cases.
    }\label{fig:slice}
\end{figure*}

\subsection{Map generation}
\label{ssec:map_generation}
Throughout this study, we consider a flat \(\Lambda\)CDM cosmology with matter density contrast:  \(\Omega_m= 0.30966\), baryon density contrast: \(\Omega_b= 0.04897\), primordial scalar spectral index: \(n_s= 0.9665\), matter fluctuation amplitude: \(\sigma_8= 0.8102\), and normalized present-day Hubble parameter in 100 Km/sec/Mpc: \(h = 0.6766\) \cite{Planck:2018}. The astrophysical parameters, normalized fraction of galactic gas in stars and corresponding power-law scaling, normalized escape fraction of ionizing photons and corresponding power-law scaling, minimum halo mass threshold for efficient star formation, and the star formation timescale taken in terms of the Hubble time, have been selected such that they produce robust cosmic dawn/epoch of reionization signal across a broad frequency range, thus enhancing detection and measurement possibilities \cite{Bonaldi:2025gpa}.  The 21-cm differential brightness temperature maps are generated on the $64^3$ regular grid across a box size of 48 Mpc. Figure \ref{fig:lightcone} shows the lightcone of the 21-cm brightness temperature from redshift 45 to 10. The top and middle panels represent the dark matter annihilation scenarios for the electron/positron channel with \(m_{\DM} = 1\)~MeV, and $\sv=1\times10^{-29}~{\cms}$. The bottom panel represents the conventional \(\Lambda\)CDM model without dark matter annihilation. The upper panel represents the case when energy injection by dark matter annihilation and deposition into the medium is inhomogeneous--- i.e., these quantities have spatial dependence. The energy injection depends on the dark matter distribution, while energy deposition efficiency mainly depends on the state of the surrounding medium. Therefore, each cell can have a different number of energetic particles produced by dark matter annihilation and outgoing flux, unlike the homogeneous scenario. The middle panel depicts the homogeneous energy injection/deposition scenario. In this case, the dark matter energy injection is calculated solely based on the global dark matter energy density, without considering its spatial distribution--- i.e., the fluctuations are neglected. Consequently, the energy injected by annihilation is evenly spread across all cells in the box. The energy deposition is determined by the total neutral hydrogen density in the simulation box and the redshift interval (\(\Delta z\)), and any remaining energy is carried over homogeneously to the subsequent redshift step. We have previously explored the latter case in detail in the context of CNNs  \cite{Nishizawa:2024bnh}. In the article, we investigated how homogeneous dark matter annihilation models can leave measurable imprints on the 21-cm signal maps that are distinguishable from scenarios without dark matter annihilation.

The distinction between inhomogeneous and homogeneous cases may not be immediately visually noticeable by looking at the lightcone. However, as previously discussed, both scenarios can exhibit distinct spatial features. In figure \ref{fig:slice}, we have presented 21-cm differential brightness temperature snaps in $xy$-plane at redshift of 12 for similar dark matter model in figure \ref{fig:lightcone}. Figure \ref{fig:slice_YInHom} represents the inhomogeneous annihilation, while figure \ref{fig:slice_NInHom} depicts the homogeneous annihilation. The Figure \ref{fig:slice_in_ho} denotes the difference between  21-cm differential brightness temperature snaps of inhomogeneous and homogeneous energy injection/deposition. As can be seen from the figure, the difference varies from about $-20$~mK to $+20$~mK. Here, we have not included the errors from sample-variance or thermal-noise from radio interferometers. Therefore, the detectability of specific dark matter models can vary based on the characteristics of the radio interferometer being used. This study focuses on the upcoming low-frequency survey of the Square Kilometer Array (SKA-Low) and examines its sensitivity to different dark matter annihilation models. In the next subsection, we discuss the sample-variance for dark matter models and system temperature noise associated with SKA-Low.

\begin{figure*}
    \begin{center}
        \subfloat[] {\includegraphics[width=0.48\textwidth]{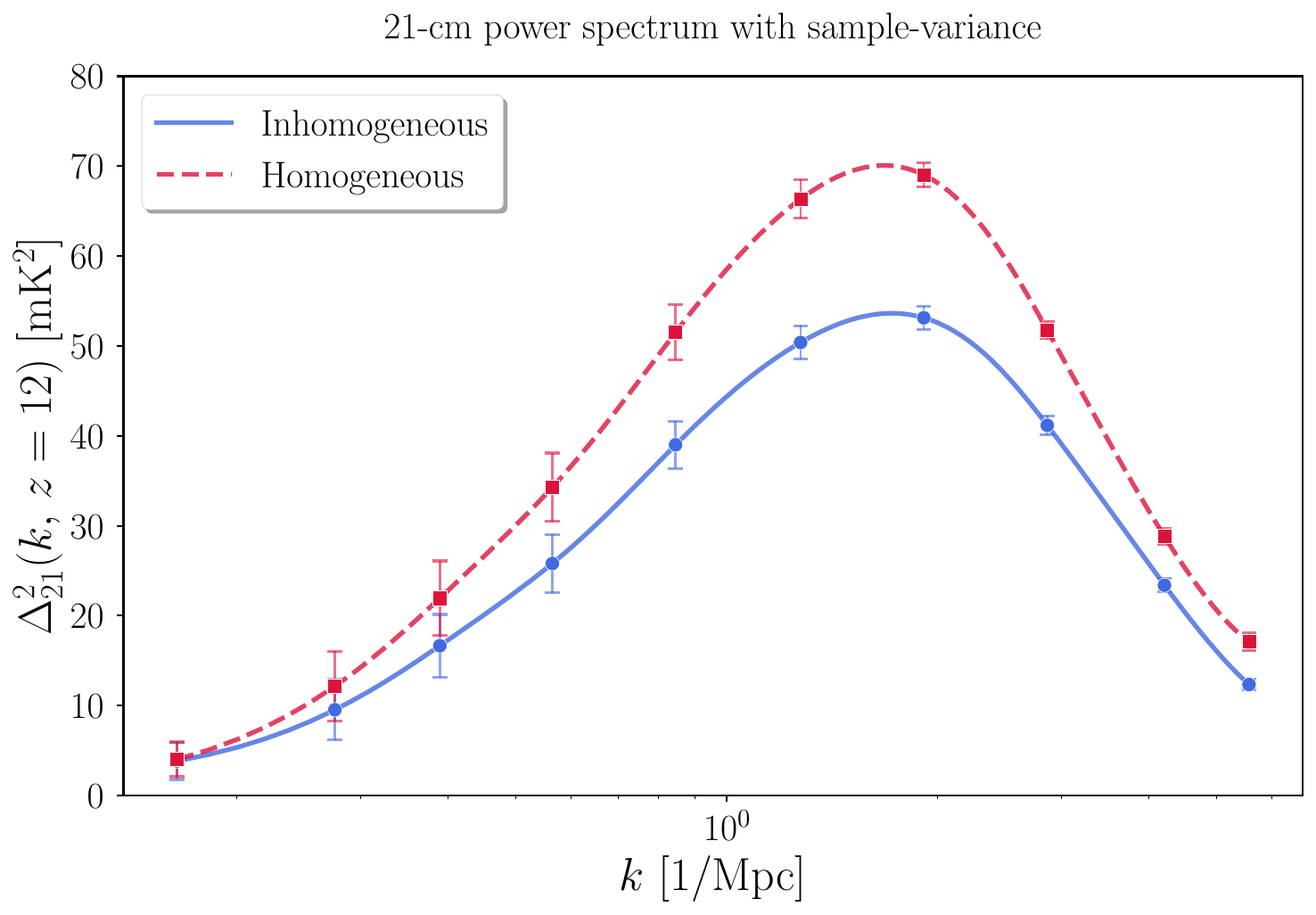}\label{fig:pk_samplevariance}}
        \subfloat[] {\includegraphics[width=0.48\textwidth]{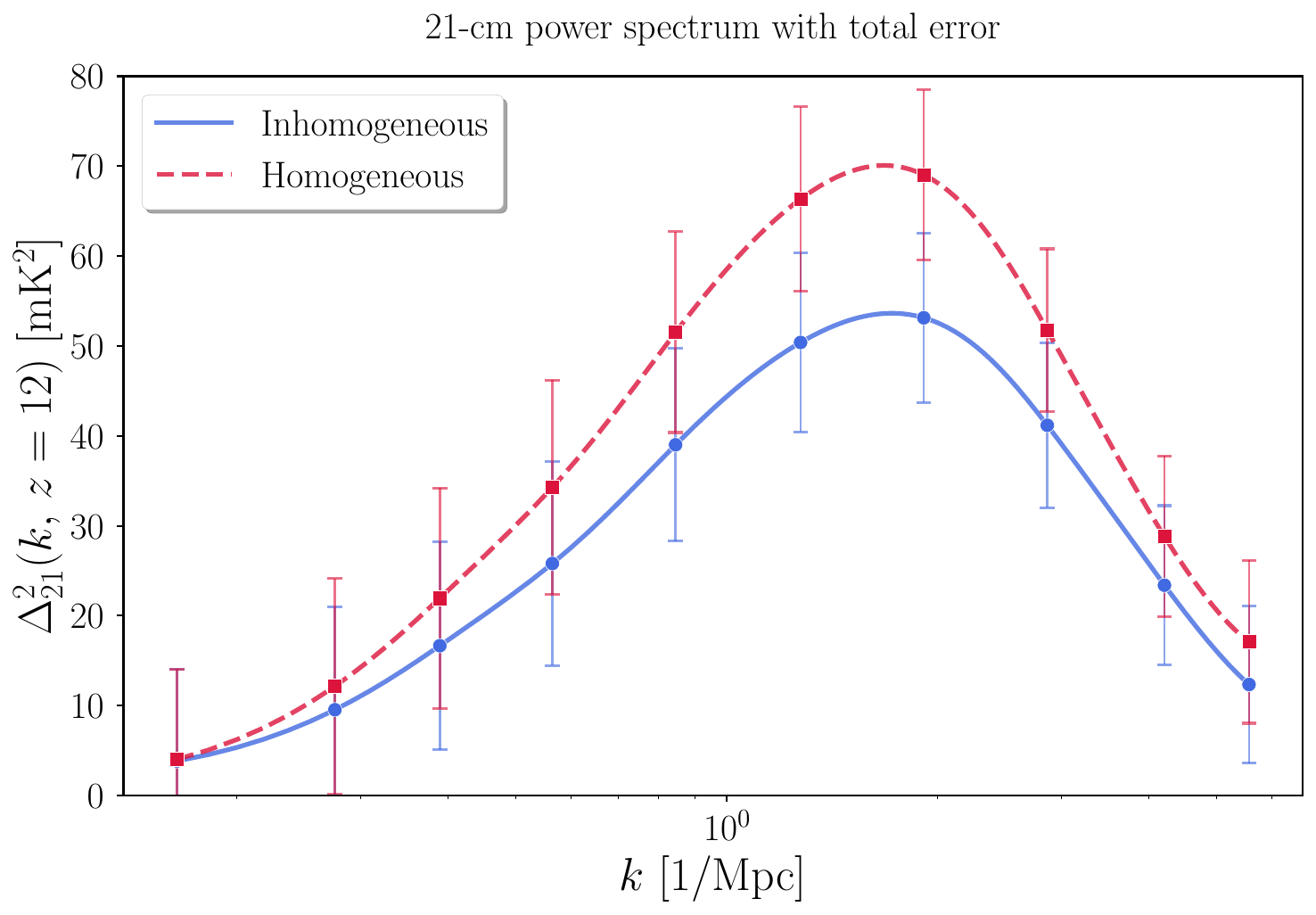}\label{fig:pk_totalerror}}
    \end{center}
    \caption{Mean 21-cm power spectrum from 1000 realization having different seeds at a redshift of 12 as a function of inverse length-scale (wavenumber-- $k$) for inhomogeneous (navy-blue solid line) and homogeneous (red dashed line) energy injection/deposition into medium for dark matter particle mass $m_{\rm DM}=1$~MeV, and cross-section $\sv=1\times10^{-29}~\cms$ annihilating to electron/positron channel. In Fig. \ref{fig:pk_samplevariance}, the error bars represent the covariance from 1000 realizations--- sample variance. In Fig. \ref{fig:pk_totalerror}, the error-bars represent the total error including sample variance and the thermal noise associated with Square Kilometre Array Low-frequency (SKA-Low) for \(f_{\rm Noise}=0.1\). 
    }\label{fig:pk_error}
\end{figure*}

\begin{figure}
    \includegraphics[width=\linewidth]{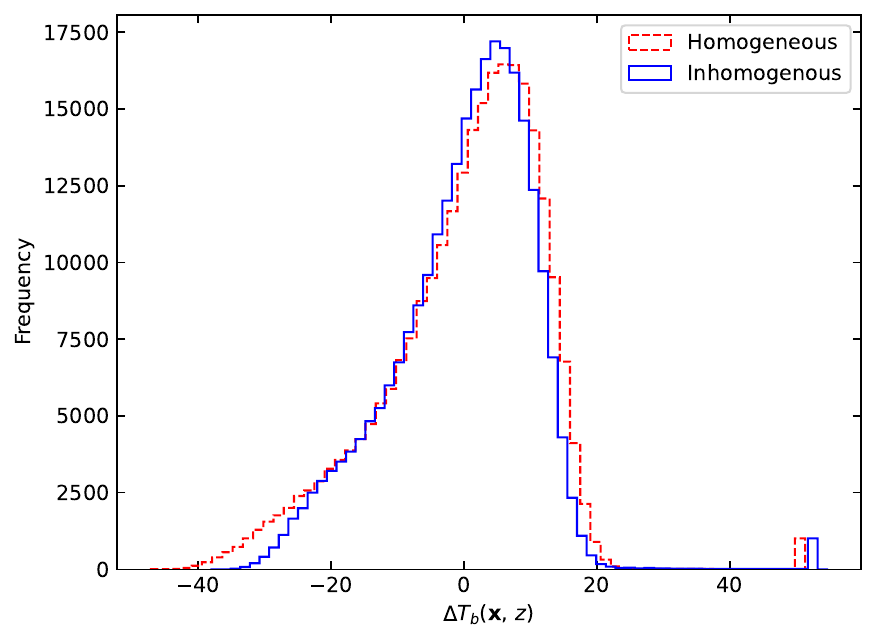}
    \caption{{Histogram for 21-cm signal at a redshift of 12 for inhomogeneous (blue solid line) and homogeneous (red dashed line) energy injection/deposition into medium for dark matter particle mass  $m_\DM=1~{\rm MeV}$, and cross-section $\sv=1\times10^{-29}~\cms$ annihilating to electron/positron channel. The histogram has been obtained after subtracting the mean from each cell. The $x$-axis represents the 21-cm signal, and the $y$-axis represents the repetition of 21-cm signal values in a particular bin.}}\label{fig:Hist}
\end{figure}
\subsection{Sample-variance and system temperature noise}
\label{ssec:noise_generation}
To accurately predict the 21-cm power spectrum, it is essential to quantify the sample-variance, which includes the uncertainty introduced by different initial conditions. We generated 1000 distinct realizations, each using a different initial seed but the same underlying dark matter model. Each realization employed a unique random seed to create the initial density field perturbations. We then calculated the power spectrum for each realization and the covariance within each \(k-\)bin. This ensemble approach directly addresses sample-variance, also known as cosmic-variance, which represents the statistical uncertainty that arises from observing a single, finite realization of the 21-cm signal. This provides a better understanding of the theoretical signal and how it varies at different scales for the specific dark matter model, allowing for more reliable and precise predictions that can be compared with observations. However, the CNN-based analysis of 21-cm maps does not require estimating the sample variance, as it is fed with all realizations. Figure \ref{fig:pk_samplevariance} represents the sample-variance in the 21-cm power spectrum for dark matter annihilation into \(\ee\) with $m_{\rm DM}=1$~MeV and $\sv=1\times10^{-29}~\cms$. The blue solid line represents the inhomogeneous case, while the red dashed line represents the homogeneous case.
{The power spectrum for the inhomogeneous case is smaller than the homogeneous case. This can be understood by considering how the distribution of energy injection/deposition affects the 21-cm signal fluctuations. In the homogeneous injection case, energy injection/deposition occurs uniformly. Consequently, the resulting 21-cm brightness temperature, $\Delta T_b$, is driven towards a more uniform state, with fluctuations primarily originating from variations in the underlying astrophysical evolution. In contrast, for inhomogeneous injection, the energy is deposited locally, strongly correlating with the highest dark matter density peaks. This creates comparatively hot and ionized ``bubbles" around these sources. As the gas temperature increases, the absorption amplitude of the 21-cm signal decreases. While these bubbles themselves exhibit lower values of $|\Delta T_b|$, the vast majority of the volume outside them experiences less heating and remains closer to the global mean temperature. This effect is clearly visible in the 21-cm signal Histogram (Fig. \ref{fig:Hist}), which shows that the distribution of $\Delta T_b$ values for the inhomogeneous case is more sharply peaked around the mean compared to the homogeneous case. In other words, the signal is more ``homogenized" in a statistical sense, with a higher probability of finding values near the average and a lower probability of finding strong outliers. This directly leads to the weaker power spectrum. The 21-cm power spectrum (eq.~\ref{eq:powersp}) is proportional to the variance of the normalized fluctuations, $\langle|\delta_{21}(\bm k,\,z)|^2\rangle_k$. When the 21-cm signal values are clustered around the mean, the magnitude of these fractional fluctuations, $|\delta_{21}|$, is systematically smaller. A smaller amplitude of $\delta_{21}$ directly translates to a lower power spectrum across all scales. As a result, the inhomogeneous scenario exhibits a weaker power spectrum compared to the homogeneous scenario.}

Figure \ref{fig:pk_totalerror} includes error bars representing the total uncertainty, which encompasses contributions from sample variance and system temperature noise in the radio interferometer instrumentation. The present analysis assumes the upcoming Square Kilometer Array (SKA) low-frequency mode survey. For simplicity, we model the noise as purely Gaussian, with zero mean, and following standard deviation in Fourier space \citep{Mao:2008ug, 2016MNRAS.459..863P},
\begin{align}
    & \sigma^2_N = \Delta^2_N\, d^2u, \label{eq:sigma_noise}
\end{align}
here, \(d^2u\) denotes the squared resolution element, which is related to the Fourier space pixel as: \(d^2l=(2\pi)^2d^2u\). Given the fact that the primary beam size depends on the area of a dish \((A_{\rm dish})\), the squared resolution element for an observed wavelength \((\lambda)\) can be expressed as: \(d^2u=A_{\rm dish}/\lambda^2\) \cite{Morales:2010}. The corresponding thermal noise power spectrum for white noise in an interferometer array can be expressed as:
\eq{ \Delta^2_N = f_{\rm Noise}\,\frac{T^2_{\rm sys}\,F^2}{B\, t_0\, n_b\,n_{\rm pol}} \label{eq:pk_noise},}
here, $f_{\rm Noise}$ denotes the fraction of the noise. We vary $f_{\rm Noise}$ from 0 to 1--- it indicates the combination of variable parameters for an interferometer to detect the specific dark matter model. The system temperature is the sum of the sky and receiver noise, and for an observing frequency, \(\nu\), it can be given approximately as \cite{2016MNRAS.459..863P},
\eq{T_{\rm sys} = 28~{\rm K}+66\,\left(\frac{\nu}{\rm 300~MHz}\right)^{-2.55}{\rm K}\,,}
$F = \lambda^2/A_{\rm dish}$ represents is the field of view. The present-day observing frequency for a 21-cm signal originating at a redshift $z$ will be $1420.4/(1+z)$~MHz. The observing frequency-width or the bandwidth of the frequency channel is estimated by the considered cell size ($\Delta\chi$),
\eq{B=\frac{\nu}{1+z}\, H\,\Delta\chi\,.}
\newcolumntype{C}{>{\centering\arraybackslash}X}
\begin{table*}
    \begin{tblr}{width=\textwidth, colspec={c | X[c] | X[c] | X[c]}, vlines={1}{0pt}}
        \hline \hline
         $f_{\rm Noise}$ & 0 & $10^{-1}$ & $10^{0}$ \\
        \hline
        $\sv=5\times10^{-28}$ [cm$^{3}$/s] & \cog{$1.00$}                                   
                                    & \cog{$1.00$}                                    
                                    & \cor{$0.55$}\\
        $\sv=1\times10^{-28}$ [cm$^{3}$/s] & \cog{$1.00$}                                  
                                    & \cog{$1.00$}                                  
                                    & \cor{$0.76$}\\
        $\sv=5\times10^{-29}$ [cm$^{3}$/s] & \cog{$1.00$}                                   
                                    & \cog{$1.00$}                                    
                                    & \cor{$0.88$} \\
        $\sv=1\times10^{-29}$ [cm$^{3}$/s] & \cog{$1.00$}                                   
                                    & \cog{$1.00$}                                    
                                    & \cor{$0.84$} \\
        $\sv=5\times10^{-30}$ [cm$^{3}$/s] & \cog{$1.00$}                                   
                                    & \cog{$1.00$}                                  
                                    & \cor{$0.72$}\\
        $\sv=1\times10^{-30}$ [cm$^{3}$/s] & \cor{$0.72$}                                  
                                    & \cor{$0.72$}                                   
                                    & \cor{$0.57$}\\
        \hline\hline
     \end{tblr}
    \caption{AUC for discriminating the homogeneous and inhomogeneous dark matter annihilation models at $z=12$ for dark matter mass $m_{\rm DM}=1$~MeV.\label{tab:auc1}}
    \vspace*{0.1cm}

    \begin{tblr}{width=\textwidth, colspec={c | X[c] | X[c] | X[c]}, vlines={1}{0pt}} 
        \hline \hline
         $f_{\rm Noise}$ & 0 & $10^{-1}$ & $10^{0}$ \\
        \hline
        $\sv=1\times10^{-27}$ [cm$^{3}$/s]
                                    & \cog{$1.00$}                                    
                                    & \cog{$1.00$}                                   
                                    & \cor{$0.82$}\\
        $\sv=5\times10^{-28}$ [cm$^{3}$/s]
                                    & \cog{$1.00$}                                    
                                    & \cog{$1.00$}                                   
                                    & \cor{$0.84$}\\
        $\sv=1\times10^{-28}$ [cm$^{3}$/s] 
                                    & \cog{$0.99$}                                    
                                    & \cog{$0.99$}                                    
                                    & \cor{$0.74$}\\
        $\sv=5\times10^{-29}$ [cm$^{3}$/s]
                                    & \cog{$0.92$}                                   
                                    & \cog{$0.91$}                                  
                                    & \cor{$0.65$}\\
        $\sv=1\times10^{-29}$ [cm$^{3}$/s] 
                                    & \cor{$0.53$}                                    
                                    & \cor{$0.55$}                                   
                                    & \cor{$0.51$}\\
        \hline\hline
    \end{tblr}
    \caption{AUC for discriminating the homogeneous and inhomogeneous dark matter annihilation models at $z=12$ for dark matter mass $m_{\rm DM}=100$~MeV. \label{tab:auc100}}
\end{table*}
In the present study, we have considered the cell size of 0.75~Mpc. Furthermore, $t_0$ is the total observation time, $n_{\rm pol}$ is the number of polarization channels, and $n_b$ is the number density of the baselines \cite{Nishizawa:2024bnh},
\begin{equation}
n_b=\frac{N_d(N_d-1)}{2\pi (u^2_{\rm max}-u^2_{\rm min})},
\end{equation}
here, $N_d$ denotes the number of stations. Further, $u_{\rm max}$ and $u_{\rm min}$ are defined in the terms of the maximum distance between antenna stations/maximum baseline ($D_{\rm max}$) and the diameter of the single antenna dish ($D_d$), respectively: $u_{\rm max}=D_{\rm max}/\lambda$, and $u_{\rm min}=D_d/\lambda$. The diameter of a single antenna dish in terms of total collecting area, $A_{\rm tot}$, can be given by,
\eq{D_d =2\, \left(\frac{A_{\rm tot}}{N_d}\,\frac{1}{\pi}\right)^{1/2}.}
In equation (\ref{eq:pk_noise}), we have defined the dimensionless power spectrum of the system thermal noise, which is utilized in the likelihood analysis with the power spectrum discussed in section \ref{ssec:model_discrimination}. We incorporate the random Gaussian noise with the variance given by equation \eqref{eq:sigma_noise} to the brightness temperature images, $\Delta T_b=\mathcal{N}(0,\sigma_N)$. 

As previously discussed, this paper assumes SKA-Low for specific computation to observe 21-cm line maps at a given redshift. We utilize the experimental setup values from the currently available SKA-Low specifications\footnote{\href{https://www.skao.int/en/explore/telescopes/ska-low}{https://www.skao.int/en/explore/telescopes/ska-low}}. Consequently, we set the parameters as follows:  $N_d=512$, $D_{\rm max}= 74~ {\rm km}$, and $A_{\rm tot} = 419,000~{\rm m^2}$. We also assume a total integration time of \(10^4\) hours, with one polarization channel and the bandwidth \(B\) calculated for each redshift using a cell size of 0.75 Mpc. 

\begin{figure}
    \includegraphics[width=\linewidth]{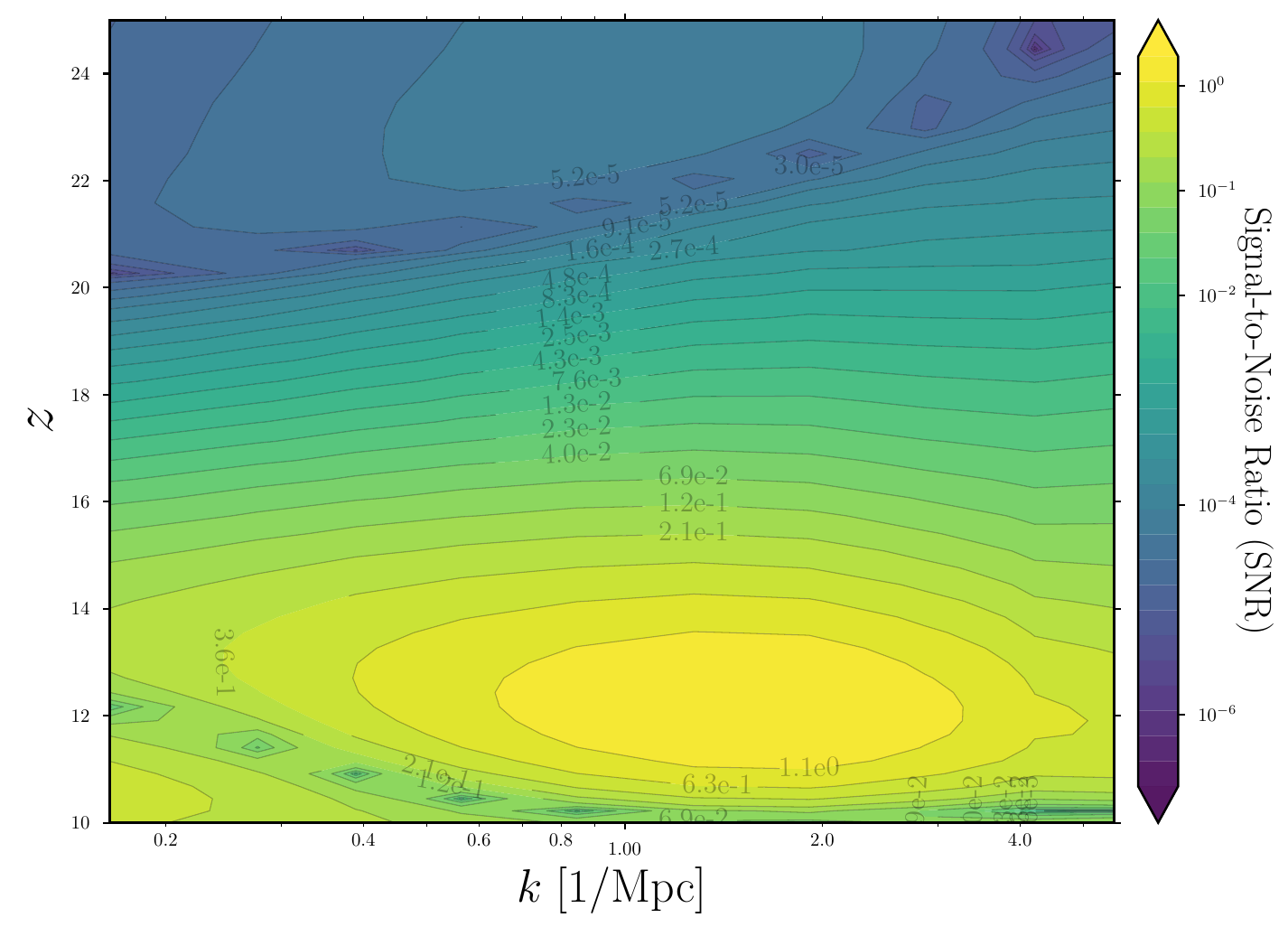}
    \caption{Signal-to-noise ratio variation with redshift ($z$) and inverse length-scale ($k$). The signal consists of the absolute difference of 21-cm power spectrum between inhomogeneous and homogeneous energy injection/deposition scenarios for dark matter particle mass $m_{\rm DM}=1$~MeV, and cross-section $\sv=1\times10^{-29}~\cms$ annihilating to electron/positron channel. The 21-cm power spectrum is calculated by averaging the power spectrum for 1000 realizations for each $k$ bin. The noise is comprised of the covariance of the difference between the 21-cm power spectra of the inhomogeneous and homogeneous energy injection/deposition scenarios, also taking 1000 realizations, along with the thermal noise associated with the Square Kilometre Array Low-frequency (SKA-Low), assuming a noise frequency of \(f_{\rm Noise} = 0.1\).}\label{fig:snr}
\end{figure}
\subsection{Model discrimination with CNN}
\label{ssec:model_discrimination}
In this subsection, we discuss the CNN process to discriminate the homogeneous and inhomogeneous dark matter annihilation models. To determine the optimal redshift for CNN analysis, we plot the signal-to-noise ratio across all values of wavenumber and redshift in Figure \ref{fig:snr}. The signal represents the absolute difference in the mean of the 21-cm power spectra for inhomogeneous and homogeneous energy injection/deposition scenarios for the fiducial dark matter annihilation model,
\eq{
{\rm signal}(k,\,z)= |\langle\Delta_{21, {\rm inh}}^2(k,\,z)\rangle_{n_{\rm rel}}-\langle\Delta_{21, {\rm hom}}^2(k,\,z)\rangle_{n_{\rm rel}}|.}
The mean of the power spectrum is calculated by averaging the power spectrum across all realizations, with \( n_{\text{rel}} = 1000 \) for each \( k \)-bin. The noise consists of two components: the sample variance from the 1000 realizations for each \( k \)-bin and the thermal noise described by equation \eqref{eq:pk_noise} for \( f_{\text{Noise}} = 0.1 \). For a given \( k \)-bin, the sample variance is calculated as: \(
|\Delta_{21, \text{inh}}^2(k, z) - \Delta_{21, \text{hom}}^2(k, z)|\), and every \( k \)-bin contains a total of 1000 elements. From the figure, it can be seen that the highest signal-to-noise ratio corresponds to the redshift $z\sim12$. Accordingly, we consider $z = 12$ as an optimal redshift for the CNN analysis of the 21-cm signal maps. A total of 25,600 images were selected from the test dataset, ensuring they were not part of the training or validation stages. We first investigate whether the inhomogeneous annihilation model could be distinguished from the homogeneous annihilation scenario. These images were evenly divided, with 12,800 representing the inhomogeneous case and 12,800 representing the homogeneous case. Discrimination accuracy was evaluated using the Area Under the Curve (AUC) of the Receiver Operating Characteristics (ROC) curve, which plots the true positive rate against the false positive rate. In this study, the inhomogeneous annihilation model was set as the positive class.  
(I) False Positive Rate (FPR): 
It measures the fraction of homogeneous annihilation scenario images that the model incorrectly labels as belonging to the inhomogeneous annihilation model. Mathematically, it is expressed as FPR = FP / (FP + TN), where: FP (false positives) is the count of homogeneous images wrongly classified as inhomogeneous; TN (true negatives) is the count of homogeneous images correctly identified as homogeneous; and the denominator, FP + TN, equals the total number of actual homogeneous images; (II) True Positive Rate (TPR): 
TPR, often called sensitivity or recall, reflects the proportion of inhomogeneous annihilation scenario images that the model accurately identifies as inhomogeneous. It is defined as TPR = TP / (TP + FN), where: TP (true positives) is the number of inhomogeneous images correctly classified as inhomogeneous, FN (false negatives) is the number of inhomogeneous images mistakenly classified as homogeneous, and the denominator, TP + FN, represents the total number of actual inhomogeneous images.
The model assigns a prediction score to each image. If this score exceeds a specified threshold--- known as the Classification Threshold, the image is classified as positive (inhomogeneous); otherwise, it is deemed negative (homogeneous). Varying this threshold adjusts the FPR and TPR values, shaping the ROC curve. The selection of the threshold is discretionary but is carefully adjusted to either maximise the gap between the ROC curve and the diagonal line in the ROC plot or to optimise the F1 score. The F1 score is computed as the harmonic mean of the TPR and PRE,
\begin{equation}
F_1 = \frac{2}{{\rm TPR}^{-1}+{\rm PRE}^{-1}},
\end{equation}
here, PRE=TP/(TP+FP) stands for the precision, representing the proportion of correctly identified positive instances among all instances predicted as positive. The value of the F1 score varies depending on the threshold selected, so we determine the threshold that yields the highest F1 score.

\begin{figure}
    \includegraphics[width=\linewidth]{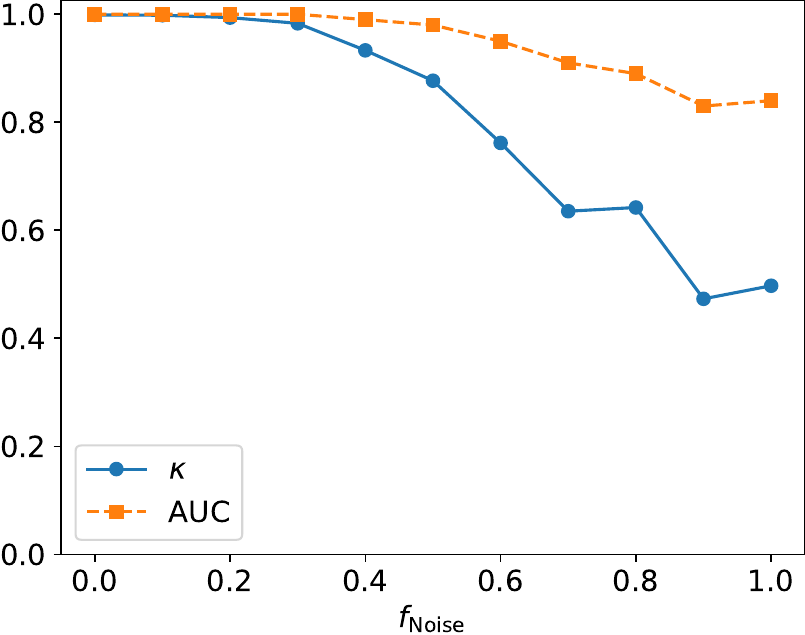}
    \caption{AUC and Cohen's $\kappa$ as function of $f_{\rm Noise}$ for fiducial dark matter annihilation model.} \label{fig:auc_kappa}
\end{figure}

We also introduce the metric to quantify the goodness of the prediction, \textit{Cohen's $\kappa$}, originally defined by \citet{cohen1960kappa} and recently applied \citep[for e.g.][]{symbolic2025astro}
\begin{equation}
    \kappa = \frac{p_o-p_e}{1-p_e},
\end{equation}
where, $p_o$ is an observed fractional agreement, and $p_e$ is expected agreement probability. Given a test sample consisting of $N_i$ inhomogeneous and $N_h$ homogeneous images, and predictions assigning $n_i$ images as inhomogeneous and $n_h$ as homogeneous, the observed agreement $p_o$ and the expected agreement $p_e$ are defined as follows
\begin{align}
    p_o &= \frac{\rm TP+TN}{N} \\
    p_e &= \left(\frac{N_i}{N}\cdot \frac{n_i}{N} \right)+
           \left(\frac{N_h}{N}\cdot \frac{n_h}{N} \right),
\end{align}
where, $N$ is the total number of samples, $N=N_i+N_h$. $\kappa$ can measure the accuracy of the prediction accounting for the imbalance of the samples; $\kappa=1$ means a perfect classification, and $\kappa=0$ is a random classification. If $\kappa$ is negative, it means the classification is dominated by some systematic errors that totally disturb the classification.

In Figure \ref{fig:auc_kappa}, we present Cohen's $\kappa$ alongside the AUC as a function of $f_{\rm Noise}$ for the case where the dark matter mass, $m_{\rm DM}=1$ MeV, and the annihilation cross-section, $ \sv = 1 \times 10^{-29}~\cms$. The figure demonstrates that the AUC shows good discrimination performance for $f_{\rm Noise}\lesssim 0.8$.

The table \ref{tab:auc1} shows the AUC values for dark matter models of mass $m_\DM=1$~MeV across various cross-sections at a redshift of 12. We consider three values for $f_{\rm Noise}$: 0, $10^{-1}$, and 1. The value $f_{\rm Noise}=0$ indicates that there is no interferometer thermal noise included. We can effectively distinguish between the inhomogeneous annihilation scenario and the homogeneous annihilation scenarios for $\sv\geq5\times10^{-30}~\cms$ at $z=12$ for $f_{\rm Noise}=0$ and $10^{-1}$. However, for $\sv<5\times10^{-30}~\cms$, the discrimination between the models becomes slightly challenging, as the AUC decreases to 0.72 for both \(f_{\text{Noise}} = 0\) and \(10^{-1}\). An AUC of about 0.5 indicates no discriminative ability, essentially equivalent to random guessing. An AUC between 0.6 and 0.7 suggests poor to fair performance, indicating limited reliability. AUC values ranging from 0.7 to 0.8 are generally considered acceptable, while those between 0.8 and 0.9 reflect nearly good performance. AUC values above 0.9 demonstrate excellent performance. The AUC value reaches a maximum of 0.88 at a value of \(\sv = 5 \times 10^{-29}~\cms\) when \(f_{\rm Noise} = 1\). {The AUC value maximises for a particular value of annihilation cross-section. The non-monotonic behaviour of AUC values can be understood by considering three regimes: (I) High Cross-Section-- A larger cross-section enhances the dark matter annihilation rate, leading to higher localized heating and ionization of the medium--- i.e., less neutral hydrogen. As a result, the mean free path for the secondary photons increases, allowing them to travel farther and reach nearby cells. This effectively smears out the energy deposition over a larger volume, enhancing the uniformity in heating and ionization of the gas within the box due to dark matter annihilation--- causing the scenario to resemble a homogeneous injection model. Consequently, it becomes more difficult to distinguish between the inhomogeneous and homogeneous annihilation models, leading to a lower AUC. (II) Intermediate Cross-Section-- As the cross-section is decreased, the energy injection becomes more localized. The mean free path of secondary photons is reduced, preventing them from propagating far and creating a clear contrast between the strongly affected regions near annihilation sites and the unaffected background. This maximizes the difference between the inhomogeneous and homogeneous scenarios, resulting in the highest AUC. (III) Very Low Cross-Section-- When the annihilation cross-section becomes too low, the overall energy injection into the medium becomes negligible. The impact on the 21-cm signal is minimal and nearly uniform, much like a very subdued version of the homogeneous case. With the distinctive, localized signatures of inhomogeneity absent, the two scenarios become difficult to distinguish again, and the value of AUC decreases. In summary, the AUC peaks at an intermediate cross-section, providing the optimal balance: sufficient energy injection to create a detectable signal, while remaining localized enough to be clearly distinguishable from a homogeneous background.}

The table \ref{tab:auc100} shows the AUC values for dark matter models of mass $m_\DM=100$~MeV across various cross-sections at a redshift of 12. As discussed earlier, an electron with an energy of 100 MeV can upscatter CMB photons to an energy of approximately 1 keV at a redshift of 12. Consequently, the mean free path for the secondary photons increases when the dark matter mass, \(m_\text{DM}\), is 100 MeV compared to when it is 1 MeV. Therefore, to enhance the local effects of dark matter annihilation, we need to increase the annihilation cross-section. As a result, the upper bound on the annihilation cross-section required to distinguish between inhomogeneous and homogeneous scenarios becomes more relaxed with a higher mass of dark matter. For this dark matter model, we can effectively distinguish between inhomogeneous and homogeneous annihilation scenarios when the cross-section, \(\sv\), is $\geq 5\times10^{-29}~\cms$ with \( f_{\text{Noise}} = 0 \) and \( 10^{-1} \). However, the AUC values start to decrease after $\sv\geq 5\times10^{-28}~\cms$ as can be seen in the last column when \( f_{\text{Noise}} = 1 \). This decline happens because a larger cross-section leads to increased energy injection into the medium, which results in higher ionization of the surrounding gas. Consequently, the medium becomes more transparent to the secondary ionizing photons, allowing them to travel to nearby cells as discussed earlier. As a result, scenarios with a higher cross-section behave similarly to cases of homogeneous energy injection. Here it is to be noted that increasing the cross-section too large can also result in an emission signal instead of an absorption signal, i.e. $T_{\rm gas}>T_{\rm CMB}$, which can be ruled out by CMB observations. When $\sv<5\times10^{-29}~\cms$, the CNN ability to discriminate becomes equivalent to random guessing (${\rm AUC\sim 0.5}$). Increasing the noise level $f_{\rm Noise}$ to 1 reduces the efficiency of the CNN in distinguishing between the inhomogeneous and homogeneous annihilation scenarios.

We also explore the dark matter annihilation to the photon-photon ($\gamgam$) channel. We explore the dark matter mass range from $10^{-2}$~MeV to GeV and scan various values of cross-section. In this scenario, we do not find any dark matter model that we can discriminate between inhomogeneous annihilation and homogeneous annihilation. The AUC values always remain below $\sim0.7$. This is due to a very large mean free path, larger than the cell size, of the injected photons. For example, even a photon having energy as low as $10^{-2}$~MeV, at a redshift of interest, can have a mean free path larger than the Hubble radius \cite{Evoli:2012}. The medium remains optically thin to photons with energies \(E_\gamma > 10^{-2}\) MeV that originated below a redshift of \(z \sim 100\). This means they can move freely in the plasma and contribute to the extragalactic X-ray and gamma-ray background. These photons can be instantly absorbed by the medium when their energy is redshifted to \(\lesssim 100\) eV \cite{Sun:2023}. For instance, if a $10^{-2}$ MeV photon is emitted at a redshift of 50 from dark matter annihilation, it can only redshift up to an energy of about 1 keV within the relevant redshift range. The free-streaming probability of high-energy photons can be defined by the ratio of the mean free path to the Hubble radius: \(\lambda_{\rm mfp} /\lambda_{H}\). The Hubble radius increases with the expansion of the Universe as \(\lambda_{H} \propto (1 + z)^{-3/2}\). Meanwhile, the mean free path for specific processes such as photoionization, Compton scattering, or pair production is inversely proportional to the gas density, given that the cross-section remains relatively constant with photon energy: \(\lambda_{\text{mfp}} \propto (1 + z)^{-3}\). Consequently, free-streaming probability or non-interacting probability becomes: \(\propto (1+z)^{-3/2}\). Hence, we can reasonably conclude that non-interacting photons at a specific cosmic epoch are even less likely to interact in a subsequent evolutionary stage. Therefore, even if photons are injected in an inhomogeneous manner, they cannot be absorbed or deposited in their respective cells. Photons with energies as low as 100 eV can have mean free paths that exceed the size of a cell, allowing them to travel to adjacent cells. Consequently, even if the number density of emitted photons increases due to a higher dark matter annihilation rate, their absorption efficiency into the plasma remains nearly unchanged because of their large mean free path. We do not find any parameter set for dark matter annihilation that would allow us to distinguish between the inhomogeneous and homogeneous annihilation models. Therefore, we do not anticipate a significant improvement in AUC values for dark matter annihilation to the \(\gamgam\) channel. {In conclusion, on-the-spot approximation is not valid for the \(\gamgam\) channel. The rationale is that the statistical uncertainty from sample variance dominates the signal--- even before considering instrumental noise. Specifically, the magnitude of the difference in the power spectra between the homogeneous and inhomogeneous injection models is significantly smaller than the associated sample variance. Therefore, the specific effect of breaking the on-the-spot approximation is not a discernible feature in the presence of this larger, inherent uncertainty.}

\section{Summary}\label{sec:summary}
This paper examines the effects of inhomogeneous and homogeneous dark matter annihilation on the 21-cm signal during the pre-reionization period, focusing on how the energy released from this annihilation process is transferred to the gas, thereby altering the measurable 21-cm differential brightness temperature. In the inhomogeneous annihilation scenario, energy injection/deposition depends on local dark density contrasts and gas properties. In the homogeneous annihilation scenario, energy deposition is spatially uniform. To generate simulated 21-cm differential brightness temperature maps, we use the {\tt DM21cm} code based on {\tt DarkHistory} and semi-numerical code {\tt 21cmFAST}. After simulating the 21-cm signal maps, we integrated the anticipated noise from the upcoming low-frequency mode survey of the Square Kilometer Array interferometer (SKA-Low) into these maps. To evaluate the required sensitivity for SKA-Low surveys aimed at distinguishing between the presence of any homogeneous heating mechanisms or any spatial heating mechanisms, we introduced an efficiency parameter denoted as \( f_{\text{Noise}} \). We multiplied this parameter by the thermal noise power spectrum of SKA-Low and varied its value from 0 to 1. To distinguish inhomogeneous dark matter annihilation from homogeneous annihilation, we take advantage of advanced machine learning techniques, specifically convolutional neural networks (CNNs). These tools help us determine whether we can distinguish between dark matter annihilating in a homogeneous manner or inhomogeneous manner based on its influence on the 21-cm signal. The results emphasize the effectiveness of CNNs in improving our ability to detect and identify the presence of any particular exotic source of energy injection into the medium with the upcoming SKA-Low data during the initial stages of structure formation in the Universe. 

In the present study, we investigate two primary channels through which dark matter can annihilate: (I) photon pairs ($\gamma\gamma$); and (II) electron-positron pairs ($e^+e^-$). We analyze a range of dark matter masses and annihilation cross-sections. Our results demonstrate that 21-cm signal-based analysis can effectively differentiate between inhomogeneous and homogeneous annihilation scenarios for an annihilation cross-section of $\sv\geq5\times10^{-30}~\text{cm}^3/\text{s}$, for a dark matter mass of $m_\DM = 1$~MeV, when the noise fraction $f_{\rm Noise} \leq 10^{-1}$ and dark matter decays to $e^+e^-$. When $\sv = 1 \times 10^{-29}~\text{cm}^3/\text{s}$, our results indicate strong discrimination capability for noise fractions $f_{\rm Noise}\lesssim 0.8$. We also consider a dark matter model with $m_\DM = 100$~MeV. For this model, we can effectively distinguish between inhomogeneous and homogeneous annihilation scenarios for $\sv\geq5\times10^{-29}~\text{cm}^3/\text{s}$. In the case when dark matter annihilates to $\gamgam$, we do not find any dark matter model for which we can discriminate between inhomogeneous and homogeneous annihilations. Here, the long mean free path of the photons erases the effects of spatial inhomogeneities. 

In conclusion, the spatial information from upcoming 21-cm observations, such as SKA-Low, will enable robust testing for the presence of any exotic sources of energy injection or deposition into the medium. Understanding the effects of spatial inhomogeneities combined with temporal information on 21-cm signal maps is crucial for making accurate predictions in the dark matter paradigm. In the future, our research will focus on enhancing these analytical techniques and applying them to real-world data from the upcoming SKA and other next-generation 21-cm observations. With improved sensitivity and resolution, these observations will open new avenues in our quest to understand the microscopic properties of dark matter.

\section*{acknowledgments}
This work is in part supported by the Hangzhou City Scientific Research Funding, No. E5BH2B0105/B015F40725006.

\appendix

\section{Full probability of the classification}
\begin{figure*}
\includegraphics[width=\linewidth]{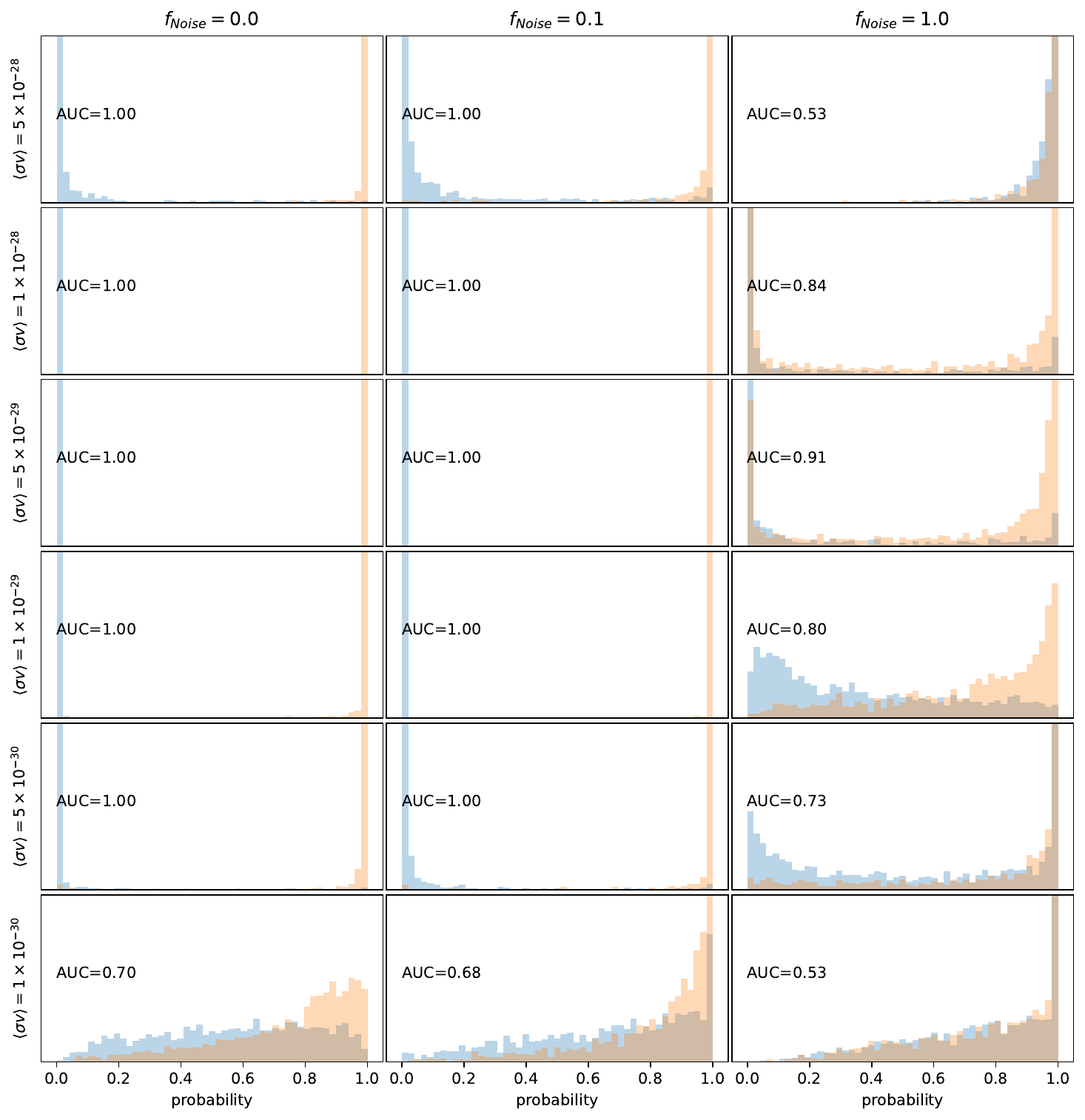}
\caption{Raw output from the CNN describing the probability of images being homogeneous (0) or inhomogeneous (1) model. Blue and orange are images from a test sample of a homogeneous and an inhomogeneous model, respectively. We also show the AUC value inset. }\label{img:raw}
\end{figure*}

Here we show the raw output of the CNN, which is the probability of the test image being a homogeneous or inhomogeneous model (Fig. \ref{img:raw}). For images classified as homogeneous (inhomogeneous) with confidence have value $p=0~(1)$, but those not in confidence, they may have values in between.

\bibliography{bibtexarxiv2}

\end{document}